\documentclass[prb,aps,twocolumn,showpacs]{revtex4}

\usepackage{graphicx,epsfig}
\usepackage{bm}
\usepackage{amsmath}
\usepackage{color}
\def\mat#1{\mathbf{#1}}

\begin{document}

\title{From thermal equilibrium to nonequilibrium quench
       dynamics: A conserving approximation for the
       interacting resonant-level}
\smallskip

\author{Yuval Vinkler-Aviv$^1$, Avraham Schiller$^{1,} \footnote{Deceased, June 22nd 2013.}$ and Frithjof B. Anders$^2$}

\affiliation{$^1$Racah Institute of Physics,
                 The Hebrew University, Jerusalem 91904,
                 Israel
						\\
				$^2$Lehrstuhl f\"ur Theoretische Physik II,
				Technische Universit\"at Dortmund,
				44221 Dortmund, Germany}

\begin{abstract}
We develop a low-order conserving approximation for the
interacting resonant-level model (IRLM), and apply it
to (i) thermal equilibrium, (ii) nonequilibrium steady
state, and (iii) nonequilibrium quench dynamics. Thermal
equilibrium is first used to carefully gauge the quality
of the approximation by comparing the results with other
well-studied methods, and finding good agreement for small
values of the interaction. We analytically show that the power-law exponent of the renormalized level width usually derived using renormalization group approaches can also be correctly obtained in our approach in the weak interaction limit.
A closed expression for the nonequilibrium steady-state current is derived and analytically and numerically evaluated. We find a negative differential conductance at large voltages, and the exponent of the power-law suppression
of the steady-state current is calculated analytically
at zero-temperature.
The response of the system to quenches is investigated for a single-lead as well as for two-lead setup at finite voltage bias at particle-hole symmetry 
using a self-consistent two-times Keldysh Green function approach, and results are presented for the time-dependent current for different bias and contact interaction strength.
\end{abstract}

\maketitle

\section{Introduction}
Describing correlated electronic systems far from thermal equilibrium is a major open problem in modern condensed-matter physics.
From the experimental side, an unprecedented control over
the microscopic parameters in nano-devices has been achieved in 
the last two decades.\cite{KastnerSET1992,NatureGoldhaberGordon1998,Kouwenhoven2000}
The simultaneous reduction of the dimensionality of devices enhances quantum
fluctuations, and correlation effects start to  dominate the physics at
low temperatures.  A large charging energy and the quantization of charge leads to new and unexpected phenomena such as lifting of the Coulomb blockade
at low temperatures.~\cite{KastnerSET1992,NatureGoldhaberGordon1998,Kouwenhoven2000}
Understanding of strong correlations in nano-devices under non-equilibrium
conditions is of fundamental  importance for their application in the 
nano-electronics of the future.

On the other hand, the description of strong electronic correlations
far from thermal equilibrium poses an enormous
theoretical challenge. At the root of the problem
lies the nonequilibrium density operator which is
not explicitly known in the presence of interactions.

In this paper, we will investigate a minimalistic model for 
quantum-transport through a nano-device: The interacting
resonant level 
model\cite{Schlottmann1980,Schlottmann1982,BohrSchmitteckert2007,RT-RG-IRLM-2,RT-RG-IRLM,%
RT-RG-IRLM-3} 
(IRLM) describes tunneling through 
a single spinless level with capacitive coupling to the leads.
This model has first been proposed an expansion of the Kondo
model around the Toulouse point:~\cite{Schlottmann1978}
the two charge states $n_d=0,1$ map on a spin 1/2
and the energy of the orbital plays the role of an external magnetic field.
The model can be solved exactly using the Bethe ansatz 
~\cite{Filyov1980}. The low temperature fixed point 
that of a non-interacting resonant level model 
whose renormalized level broadening can be calculated
using a renormalization group approach.
~\cite{Schlottmann1978,Schlottmann1980,Schlottmann1982}
Since its equilibrium properties are well understood, this
model can serve as ideal non-trivial test for conserving approaches
which are applicable  to the equilibrium as well as to the non-equilibrium
regime.

Recently, a scattering states Bethe ansatz approach
for the calculation of steady state currents\cite{MethaAndrei2006}
has been proposed triggering
a lot of investigation on the non-equilibrium dynamics in this model.
\cite{BoulatSaleurIRLM2008,BoulatSaleurSchmitteckert2008}
A negative differential conductance for large bias has been
found\cite{BoulatSaleurSchmitteckert2008} in a study combining TD-DMRG\cite{Schmitteckert2004,DaleyKollathSchollwoeckVidal2004,Schollwoeck2011}
and Bethe-ansatz results at the duality point\cite{SchillerAndrei2007-I} 
based on a power law decay of the current.
This surprising result has been linked to a frequency dependent
renormalization of the charge fluctuation scale by replacing 
the frequency with the applied bias in 
a perturbative renormalization group (RG) approach.~\cite{BordaZawa2010} 
Similar findings have been also obtained using 
functional-RG approaches.~\cite{RT-RG-IRLM-2,RT-RG-IRLM}

In this paper we will show that the negative differential conductance found
in the state of the art numerics~\cite{BoulatSaleurSchmitteckert2008}
or in  perturbative RG approaches~\cite{BordaZawa2010,RT-RG-IRLM-2,RT-RG-IRLM}
can also  be obtained employing the lowest
order conserving approximation.~\cite{Baym62,Dubi2013}
We present a closed analytical solution of the self-consistency  equation 
of a conserving approximation 
for $T=0$ and for large temperature in equilibrium which agrees remarkably 
well with the perturbative RG solution in the weak coupling 
limit. We will analytically calculate exponent of renormalized level width
which agrees perfectly with a recent functional renormalization group (fRG) approach\cite{Kennes2012} based on the same Hartree diagram.

After establishing the accuracy
of our method, we extend our  Kadanoff-Baym-Keldysh~\cite{kadanoffBook62,Keldysh65}  approach
to the steady state non-equilibrium and present results for the $I-V$ curves.
In the linear response regime, universality of the differential conductance
is reproduced which is a consequence of the  universal 
local Fermi liquid fixed point of the model.~\cite{KrishWilWilson80a,BullaCostiPruschke2008}

Using the full time dependency of the non-equilibrium two-times Green functions\cite{Baym62}
we  calculate the real-time evolution of the current 
after switching on the tunnelling term at time $t=0$. We can show that our conserving approximation
always approaches the steady-state limit for long times. We analyze our numerical solution for the interacting problem by comparing it to the exact analytical expression of the time-dependent current for the non-interacting case. $I(t)$ can be qualitatively understood by replacing the bare charge fluctuation scale $\bar{\Gamma}_0$ in the non-interacting case by the steady-state renormalized value $\bar{\Gamma}_{\rm eff}$ depending on the interaction strength. Quantitatively, however, we observe significant differences in the short and intermediate time behavior:
the time-dependence of the charge fluctuation scale $\bar{\Gamma}_{\rm eff}(t)$ influences not only the initial slope of the current but also increases amplitude of the current oscillations at finite voltages while simultaneously decreasing the decay rate of these oscillations with increasing interaction strength.

This increase of the current oscillation amplitude has already been
previously observed in an recent elaborate functional RG and a real-time RG study\cite{Kennes2012} away from the particle-hole symmetry point. In this paper  we demonstrate that (i) increase of the current oscillation amplitude 
is generic feature prevailing in the particle-hole symmetric case and (ii) 
a simple conserving Hartree approximation is sufficient to derive 
the power-law renormalization of the charge fluctuation scale as well as (iii)
the power-law suppression of the steady-state current at large voltage.

\section{The model and conserving approximation}

\subsection{The interacting resonant-level model}

Our model of interest -- the IRLM -- describes a single
spinless level $d^{\dagger}$ which is both hybridized
with one or more spinless bands of electrons, and subject
to a contact interaction with the bands. This is the
most elementary extension of the standard non-interacting
resonant-level model to account for interactions that
take place in a tunnel junction. The model has a long
history that dates back to the 1970's, when it was
proposed as a minimal model for valence-fluctuating
systems. In recent years it has regained considerable
interest as a generic model for the combined study of
interactions and nonequilibrium conditions.

Formally, the $M$-channel IRLM is defined by the
Hamiltonian
\begin{eqnarray}
	\mathcal{H} &=&
	\sum_k\sum_{\alpha=1}^{M}
	\left(\epsilon_k-\mu_\alpha\right)
	c^{\dagger}_{\alpha k} c_{\alpha k}
	\nonumber \\ &&
	+\sum_{\alpha=1}^{M}\frac{\gamma_{\alpha}}{\sqrt{N}}\sum_{k}
	\left(c^{\dagger}_{\alpha k}d + {\rm h.c.}
	\right)	
	+	\epsilon_d d^{\dagger}d 
	\nonumber \\ &&
	+ \frac{U}{N}
	\left(d^{\dagger}d-\frac{1}{2}\right)
	\sum_{k,q}\sum_{\alpha=1}^{M}
	:\!c^{\dagger}_{\alpha k}c_{\alpha q}\!:,
	\label{H_general}
\end{eqnarray}
where $c^{\dagger}_{\alpha k}$ creates a conduction electron with energy 
$\epsilon_k$ in channel $\alpha$, and $d^\dagger$ creates an electron in
a single localized orbital with energy $\epsilon_d$ modelling  the nano-device.
Here, $\gamma_\alpha$ is the hopping matrix element to the channel $\alpha$, which has a chemical potential $\mu_\alpha$ so that a current can be driven through such a junction. $U$ labels the contact interaction, assumed to be identical for all bands and stemming from the capacitative coupling between a localized electron and the surrounding electron gas.
$N$ is the number of lattice sites (i.e., the number
of $k$ points) in each band,
and $:\!c^{\dagger}_{k n} c^{}_{k' n}\!: =
c^{\dagger}_{k n} c^{}_{k' n} - \delta_{k, k'}
\theta( -\epsilon_k )$ stands for normal ordering
with respect to the filled Fermi sea.
For simplicity, we assume
particle-hole symmetric bands with identical dispersion for all channels. Written in this form, resonance condition corresponds
to $\epsilon_d = 0$, when the model is manifestly
particle-hole symmetric for $\mu_\alpha=0$.

\subsection{Conserving approximation}
\label{sec:conserving-approximation}
The approximation we shall employ in this paper follows the approach introduced by Baym~\cite{Baym62} for treating the Coulomb gas. The self-energies $\Sigma$ are defined as functional derivatives of a generating functional $\Phi$, which is written using the fully dressed Green functions
\begin{equation}
	\Sigma_{AB} = \frac{\delta \Phi}{\delta G_{AB}},
\end{equation}
where $A$ and $B$ are the degrees of freedom of the system to which the self energy pertains. The diagrammatic representation of $\Phi$ resembles the perturbative expansion for the ground state energy of the system. The Green functions are then calculated by solving the self-consistency equation derived from this definition of the self-energies. This approximation is consistent with microscopic conservation laws, and guarantees correlation functions that respect these laws.

\begin{figure}[b]
\centerline{
\includegraphics[width=85mm]{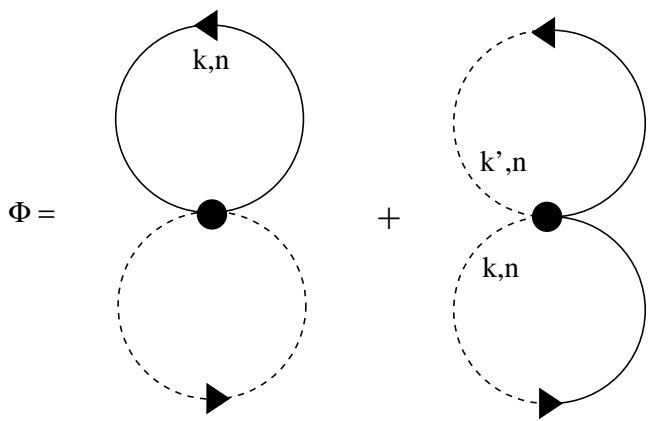}
}\vspace{0pt}
\caption{The generating functional $\Phi$ defining
         our conserving approximation. The conduction electrons are depicted by continuous lines labeled by the momentum $k$ and channel number $n$, the level's degree of freedom is represented by dashed lines and the value of each vertex is $U$. A summation over all different momenta $k$, $k'$ and channels $n$ is assumed.
        }
\label{Fig:fig1}
\end{figure}

The generating functional for the model at hand, defining our conserving approximation, is portrayed in Fig.~(\ref{Fig:fig1}). It is perturbative in the contact interaction $U$, and contains the leading-order diagrams describing that interaction. The quality of the approximation is controlled by the small parameter $\rho_0 U$, where $\rho_0$ is the density of states at the Fermi energy, limiting our results to small values of the interaction with respect to the bandwidth. However, previous works have shown that the interacting resonant level model has a duality between strong and weak values of the contact interaction for the case of two screening channels $M=2$, 
and derived an analytical mapping between the strongly and weakly interacting models, which is applicable also far from thermal equilibrium.~\cite{SchillerAndrei2007-I,MethaAndrei2006} Building on the these results, our treatment of the model for small values of $\rho_0 U$ can be extended to strong values $\rho_0 U'$ defined by the mapping
\begin{equation}
	\pi\rho_0 U ' = \frac{4}{\pi\rho_0 U}.
\end{equation}
This mapping allows us to compare our results with studies of this model that use methods that are geared toward strong interaction such as the hybrid td-NRG/td-DMRG.

A word is in order with respect to the perturbative RG approach of Ref.~(\onlinecite{BordaZawa2010}).
Although our results are very similar to the predictions of Borda et al, there are major technical differences
between the approaches. We only consider contributions linear in the interaction strength 
while Borda et al include second order loop corrections. While Ref.~(\  \onlinecite{BordaZawa2010}) is
a perturbative RG calculation strictly speaking only well justified in equilibrium and in the limit of large number $M$ of screening channels, we consider a full
self-consistent Kadanoff-Baym-Keldysh approach which 
holds for any number of screening channels, arbitrary temperature and voltage as long as the dimensionless
coupling constant $g=\rho_0 U$ remains small.  In our approach, see details below, the voltage dependence
occurs naturally while in  Ref.~(\  \onlinecite{BordaZawa2010})  the $eV/2$ is substituted by hand  for
the frequency $\omega$.

\section{Thermal equilibrium}
\label{sec:termal_eq}

To set the stage for our non-equilibrium calculations,
we begin with a discussion of the thermal equilibrium
and set $\mu_\alpha=0$. Only the binding linear combination
\begin{eqnarray}
\gamma \tilde c_{k} &=& \sum_\alpha \gamma_\alpha c^\dagger_{\alpha k},
\end{eqnarray}
hybridizes with the $d$-orbital where $\gamma^2 = \sum_\alpha \gamma_\alpha^2$.
For degenerate bands, we perform a unitary transformation to $c_{1k}=\tilde c_k$ and
label are orthogonal linear combinations as $n=2,\dots, M$. Consequently,
we arrive at the Hamiltonian
\begin{align}
{\cal H} &= \sum_{n = 1}^{M} \sum_k
                 \epsilon_k c^{\dagger}_{k n} c^{}_{k n}
          + \frac{\gamma}{\sqrt{N}}
            \sum_k
                 \left \{
                          d^{\dagger} c^{}_{k, n = 1}
                          + {\rm H.c.}
                 \right \}
\nonumber \\
         & + \epsilon_d d^{\dagger} d
           + \frac{U}{N}
             \left(
                    d^{\dagger} d - \frac{1}{2}
             \right)
             \sum_{n = 1}^{M}
             \sum_{k, k'}
                  :\!c^{\dagger}_{k n} c^{}_{k' n} \!: .
	\label{H-equil}
\end{align}
Even though, the multi-channel version of the model contains
interesting physics\cite{BordaSchillerZawadowski2008}, we restrict
ourselves to a single channel (M=1.) Therefore, we drop the index $n$ 
in the following.

In the absence of the contact interaction $U$,the
non-interacting resonant-level model describes a
simple resonance of width $\Gamma_0 = \pi\rho_0 \gamma^2$,
where $\rho_0$ is the conduction-electron density
of states at the Fermi level and is exactly solvable.
It has been established\cite{Schlottmann1980}
that the low-energy fixed point of the  IRLM is equivalent
to its non-interacting counterpart,
both describing a phase-shifted Fermi liquid\cite{Wilson75}.

The contact interaction dresses the Green function of the 
level, and its spectral function
\begin{equation}
	\rho_d(\epsilon) = -\frac{1}{\pi}
			{\rm Im}\{G_{dd^{\dagger}}(\epsilon+i\eta)\},
\end{equation}
where $G_{dd^{\dagger}}(\epsilon+i\eta)$ is the retarded
Green function pertaining to the level's degree of freedom. In the low frequency spectrum, it is characterized by a width $\Gamma_{\rm eff}$ which is the
effective tunneling rate. The calculation of this
observable is our main goal in this section.

We begin by noting that only two degrees of freedom
appear both in the tunneling term and the interaction
term of the IRLM Hamiltonian: the level $d^{\dagger}$
and the local conduction electron at the origin,
\begin{equation}
\psi^{\dagger} = \frac{1}{\sqrt{N}}\sum_k c^{\dagger}_k.
\end{equation}
It is therefore convenient to define a $2\!\times\!2$
matrix Green function
\begin{equation}
 \mat{G}(z) = \left[
               \begin{array}{cc}
                     G_{\psi \psi^{\dagger}}(z) &
                     G_{d \psi^{\dagger}}(z) \\ \\
                     G_{\psi d^{\dagger}}(z) &
                     G_{d d^{\dagger}}(z)
               \end{array}
       \right] ,
\end{equation}
where
\begin{equation}
G_{AB}(z) = \ll\! A, B\! \gg_z
\label{def_of_G_AB}
\end{equation}
is the correlator of the fermionic operators $A$ and
$B$ in energy domain. The dressed Green function is
given in terms of the self energy matrix by
\begin{eqnarray}
	\mat{G}(z) &=&
   \left[
      \begin{array}{cc}
            g^{-1}_0(z) - \Sigma_{\psi \psi^{\dagger}}(z) &
            -\Sigma_{d \psi^{\dagger}}(z) \\ \\
            -\Sigma_{\psi d^{\dagger}}(z) &
            z - \epsilon_d - \Sigma_{d d^{\dagger}}(z)
      \end{array}
   \right]^{-1} ,
\label{G-dressed}
\end{eqnarray}
where $\Sigma_{\psi \psi^{\dagger}}(z)$,
$\Sigma_{d \psi^{\dagger}}(z)$,
$\Sigma_{\psi d^{\dagger}}(z)$, and
$\Sigma_{d d^{\dagger}}(z)$ are the self-energy
components and
\begin{equation}
	g_0(z) = \frac{1}{\sqrt{N}}\sum_k \frac{1}{z - \epsilon_k}
\label{g_bare}
\end{equation}
is the bare Green function pertaining to $\psi$ for $\gamma=U=0$.

In order to derive closed analytical expressions, we assume a Lorentzian
density of states with half width $D$ for the
conduction electrons, such that
\begin{equation}
	g_0(z) = \pi\rho_0
	\frac{1}{ z/D + i{\rm sgn} \left(
                            {\rm Im}\{z\}
                \right)},
\end{equation}
where $\rho(\epsilon)={\rm Im}\{g_0(\epsilon-i\eta)\}/\pi$ denotes the density of states of the conduction electrons, and we set $\rho(0)=\rho_0$ to be the density of states at the Fermi energy. Throughout most of this paper we shall assume that the bandwidth $2D$ is the largest energy scale in the system, and under such conditions the specific structure of the spectral function has no qualitative effects on our results. The only time where we will allow another energy scale to be comparable to $D$ will be when we will consider the system under large voltage bias.

Next we turn to the self-energies derived from the
generating functional of Fig.~(\ref{Fig:fig1}).
Within the self-consistent Hartree-approximation of
Fig.~(\ref{Fig:fig1}) the three self-energies are static
and independent of energy:
\begin{align}
& \Sigma_{d d^{\dagger}} =
        U \langle :\!\psi^{\dagger} \psi\!: \rangle ,
	\label{eq:sigma-dd}
\\ 
& \Sigma_{\psi \psi^{\dagger}} = 
        U \langle d^{\dagger}d-\frac{1}{2}\rangle ,
\\
& \Sigma_{\psi d^{\dagger}} = \Sigma_{d \psi^{\dagger}}^*
     = \gamma -U\langle \psi^{\dagger} d\rangle .
     \label{eq:sigma-psi-d}
\end{align}
As such, they have a natural interpretation as energy shifts in an effective bi-linear Hamiltonian $H\to H_{\rm eff}$ approximating H
of Eq.~(\ref{H-equil}): $\Sigma_{d d^{\dagger}}$
renormalizes the level energy $\epsilon_d$,
$\Sigma_{\psi \psi^{\dagger}}$ corresponds to local
potential scattering of the conduction electrons, and
$\Sigma_{\psi d^{\dagger}} = \Sigma_{d \psi^{\dagger}}^*$
renormalizes the hopping amplitude between the lead and
the level. Therefore, the Green function of Eq.~(\ref{G-dressed}) retains its non-interacting form, only with renormalized
couplings consistent with the unchanged low energy fixed point.

In the following we shall focus on resonance conditions,
i.e., $\epsilon_d = 0$, and assume a symmetric band with
$\rho(\epsilon)=\rho(-\epsilon)$. Under these conditions
the IRLM Hamiltonian is invariant under the particle-hole
transformation $c_k \to c^{\dagger}_{-k}$,
$d \to -d^{\dagger}$, which fixes the expectation values
\begin{align}
& \langle d^{\dagger}d \rangle = \frac{1}{2} ,
\\
& \langle :\!\psi^{\dagger} \psi\!:\rangle = 0 .
\end{align}
Consequently, $\Sigma_{\psi \psi^{\dagger}}$ and
$\Sigma_{d d^{\dagger}}$ are both zero, leaving only
the off-diagonal self-energy components. The dressed
Green function of the level is then given by
\begin{equation}
G_{d d^{\dagger}}(z) =
        \frac{1}{ z - g_0(z) |\gamma_{\rm eff}|^2 } ,
\end{equation}
where we have defined $\gamma_{\rm eff} = \gamma + U \langle
\psi^{\dagger} d \rangle$. The role of the
interaction in this approximation is now transparent:
it renormalizes the resonance width from its bare
value $\Gamma_0$ to
\begin{equation}
\Gamma_{\rm eff} = \pi \rho_0 |\gamma_{\rm eff}|^2 .
\end{equation}
Our remaining task is to compute $\gamma_{\rm eff}$
and thus $\Gamma_{\rm eff}$ in order to fully
determine $G(z)$.

After substituting the off-diagonal matrix element of 
$\mat{G}$,
\begin{equation}
G_{d \psi^{\dagger}}(z) = \gamma^*_{\rm eff}
           \frac{ g_0(z) }{ z - g_0(z)|\gamma_{\rm eff}|^2 } ,
\label{equilibrium_G_dpsi}
\end{equation}
in the self-consistency equation~(\ref{eq:sigma-psi-d})
\begin{equation}
\gamma_{\rm eff} - \gamma
       = -U \langle \psi^{\dagger} d \rangle
       = -\frac{U}{\beta}
          \sum_{n} G_{d \psi^{\dagger}}(i\omega_n) ,
\label{Matsubara}
\end{equation}
the summation over the  Matsubara
frequencies $\omega_n = \pi(2n + 1)/\beta$
can be carried out analytically
\begin{align}
-\frac{1}{\beta}
          \sum_n G_{d \psi^{\dagger}}(i\omega_n) &=&
          \frac{ \rho_0 \gamma_{\rm eff} }{x}
               \biggl [
                        \psi\left(
                                    \frac{1}{2} + (1 + x)
                                    \frac{\beta D}{4\pi}
                             \right)
\nonumber \\
                        &&- \psi \left(
                                        \frac{1}{2} +
                                        (1 - x)
                                        \frac{\beta D}
                                             {4\pi}
                                \right)
               \biggr ] ,
\end{align}
for a Lorentzian density of states, where $x$ equals $\sqrt{ 1 - 4\Gamma_{\rm eff}/D }$ and $\psi(z)$ is the digamma function.~\cite{AS-psi} Here $\beta=1/T$ is the reciprocal temperature.

Writing the self-consistency equation directly for
$\Gamma_{\rm eff}$, we finally get
\begin{equation}
\Gamma_{\rm eff} =
        \frac{\Gamma_0}
             { \left[
                       1 - \rho_0 U
                           \Lambda(\Gamma_{\rm eff})
               \right]^2 }
\label{SC-equilibrium}
\end{equation}
with
\begin{align}
\Lambda(\Gamma_{\rm eff}) =
        \frac{1}{x} &
        \left[
               \psi \left(
                            \frac{1}{2} + (1 + x)
                                 \frac{\beta D}{4\pi}
                    \right)
        \right.
\nonumber \\
&        \left.
           \;\; -
               \psi \left(
                            \frac{1}{2} + (1 - x)
                                 \frac{\beta D}{4\pi}
                    \right)
        \right] .
	\label{eq:gamma-eff}
\end{align}

Equation~(\ref{SC-equilibrium}) constitutes the
central result of this section, as its solution
yields the renormalized hybridization width
$\Gamma_{\rm eff}$, and with it the full matrix
Green function $\mat{G}(z)$. Generally, one must resort
to numerics to solve for $\Gamma_{\rm eff}$, a
task we shall undertake below. But first, let us
consider certain limits where analytical results
can be obtained.

\subsection{Weak coupling, zero temperature}
\label{sec:zero_temp_eq}

Consider first the zero-temperature limit,
$T \to 0$, when each of the digamma functions in
Eq.~(\ref{SC-equilibrium}) reduces to a log by
virtue of the asymptotic expansion~\cite{AS-psi}
\begin{equation}
	\psi(z) = \ln(z) + O(z^{-1}) .
	\label{eq:digamma-expansion}
\end{equation}
Since we are interested in wide-band limit, i.e., $D \gg \max \{ t, U, \Gamma_0, \Gamma_{\rm eff} \}$,
one can approximate $x \simeq 1 - 2\Gamma_{\rm eff}/D$.
These two simplifications lead to the compact expression
\begin{equation}
\Lambda( \Gamma_{\rm eff} ) \simeq	
         \ln \left(
                     \frac{D}{\Gamma_{\rm eff}}
             \right) ,
\label{Lambda-as-log}
\end{equation}
resulting in
\begin{equation}
\Gamma_{\rm eff} \simeq
       \frac{\Gamma_0}
            {\left[
                     1 - \rho_0U\ln(D/\Gamma_{\rm eff})
             \right]^2} .
\label{SC-T=0-via-log}
\end{equation}
Here we have omitted terms of order
$\Gamma_{\rm eff}/D$ in writing
Eq.~(\ref{Lambda-as-log}). If we further assume
sufficiently weak coupling such that
$\rho_0 U \ln(D/\Gamma_{\rm eff}) \ll 1$ (a condition
whose domain of validity we examine below), then
$1 - \rho_0 U \ln( D/\Gamma_{\rm eff} )$ is well
approximated by $(D/\Gamma_{\rm eff})^{-\rho_0 U}$,
which, when inserted into Eq.~(\ref{SC-T=0-via-log}),
yields the power-law behavior
\begin{equation}
\Gamma_{\rm eff} \simeq
       D \left(
                 \frac{\Gamma_0}{D}
         \right)^{1/(1 + 2\rho_0 U)},
\label{G_eff-RG-II}
\end{equation}
where the perturbative RG analysis of the model yields the power-law behavior~\cite{Schlottmann1982}
\begin{equation}
	\Gamma_{\rm eff} = D\left(
                 \frac{\Gamma_0}{D}
         \right)^{\frac{1}{1+2\rho_0U+(\rho_0 U)^2}}.
\end{equation}
Thus, the self-consistent approximation
coincides with leading order perturbative RG provided
$\rho_0 U \ln(D/\Gamma_{\rm eff}) \ll 1$.

The difficulty with determining the range of
validity of the condition above is that it
involves the renormalized width $\Gamma_{\rm eff}$,
which {\em a-priori} is not known. Still, one can
check its consistency with Eq.~(\ref{G_eff-RG-II})
by adopting the latter expression for
$\Gamma_{\rm eff}$, which gives
\begin{equation}
1 \gg \frac{\rho_0 U}{1 + 2 \rho_0 U}
           \ln (D/\Gamma_0) .
\label{WC-condition-I}
\end{equation}
Alternatively, Eq.~(\ref{WC-condition-I}) can be
recast in the form $\ln(D/T_U) \gg \ln(D/\Gamma_0)$,
where
\begin{equation}
T_U = D e^{-(1 + 2 \rho_0 U)/\rho_0 U}
\end{equation}
is a new energy scale that depends exclusively on
$\rho_0 U$ and $D$. As $U \to 0$ then $T_U \to 0$,
extending the range of validity of the power-law
form of $\Gamma_{\rm eff}$ to all values of
$\Gamma_0$. However, as $U$ is increased then $T_U$
increases, restricting the power-law form to the region
where $\ln(D/T_U) \gg \ln(D/\Gamma_0)$. We emphasize
that the logarithmic nature of this latter condition
makes it far more stringent than the simpler
restriction $\Gamma_0 \gg T_U$. Below we validate
this picture numerically.

\subsection{Weak coupling, finite temperature}
\label{Sec:Eq-finite-T}

Next we proceed to finite temperature $T$. Since
$x \simeq 1 - 2\Gamma_{\rm eff}/D$ still holds,
we expand Eq.~(\ref{eq:gamma-eff}) to
\begin{equation}
\Lambda( \Gamma_{\rm eff} ) \simeq
       \left[
               \psi \left(
                            \frac{\beta D}{2\pi}
                    \right)
             - \psi \left(
                            \frac{1}{2} +
                            \frac{\beta \Gamma_{\rm eff}}
                                 {2\pi}
                    \right)
       \right] ,
\label{Lambda-finite-T}
\end{equation}
where again we have omitted terms of order
$\Gamma_{\rm eff}/D$ and $T/D$. The role of a
temperature is now clear. When
$\Gamma_{\rm eff} \gg T$, each of the digamma
functions in Eq.~(\ref{Lambda-finite-T}) has a large
argument, justifying their asymptotic expansion
in Eq.~(\ref{eq:digamma-expansion}). Consequently, Eq.~(\ref{Lambda-as-log}) is recovered, up to corrections of order
$T/D$ and $T/\Gamma_{\rm eff}$.

As $T$ exceeds $\Gamma_{\rm eff}$, the argument of
the  second digamma function in Eq.~(\ref{Lambda-finite-T}) 
approaches $1/2$, and  $\psi(1/2) = -\gamma
- 2\ln(2)$ where $\gamma = 0.57721\ldots$ is Euler's constant.
Therefore,  Eq.~(\ref{Lambda-as-log})
is replaced by
\begin{equation}
\Lambda( T ) \simeq	
         \ln \left(
                     \frac{2 e^\gamma D}{\pi T}
             \right)
       \simeq \ln \left(
                     1.13 \frac{D}{T}
             \right) ,
\end{equation}
resulting in
\begin{equation}
\Gamma_{\rm eff} \simeq
       \frac{\Gamma_0}
            {\left[
                     1 - \rho_0 U \ln( 1.13 D/T )
             \right]^2} .
\end{equation}
In agreement with the perturbative RG, the temperature
$T$ is seen to replace $\Gamma_{\rm eff}$ as the
low-energy cutoff if $T > \Gamma_{\rm eff}$. As before,
we may approximate $1 - \rho_0 U \ln( 1.13 D/T )$
with $( 1.13 D/T )^{-\rho_0 U}$ if
$\rho_0 U \ln( 1.13 D/T ) \ll 1$, reproducing
the perturbative RG result
\begin{equation}
\Gamma_{\rm eff}(T)
        \simeq \Gamma_0
               \left(1.13
                       \frac{D}{T}
               \right)^{2 \rho_0 U},
\label{SC-G-vs-T}
\end{equation}
in lowest order in the dimensionless coupling constant $\rho_0U$.

\subsection{Breaking particle-hole symmetry: nonzero $\epsilon_d$}

So far, we have focused exclusively on
$\epsilon_d = 0$. For completeness, we briefly
address in this section the general off-resonance
case where $\epsilon_d \neq 0$. As emphasized above,
a nonzero $\epsilon_d$ breaks the particle-hole
symmetry of the IRLM Hamiltonian, rendering the
two diagonal self-energies $\Sigma_{d d^{\dagger}}$
and $\Sigma_{\psi \psi^{\dagger}}$ nonzero. Therefore, a
complete treatment of the off-resonance case
requires therefore a coordinated self-consistent
solution of all three parameters $\gamma_{\rm eff}$,
$\Sigma_{d d^{\dagger}}$, and
$\Sigma_{\psi \psi^{\dagger}}$. As our interest
lies in the renormalized hybridization width
$\Gamma_{\rm eff}$, we shall not attempt such a
complete treatment. Rather, we shall adopt the
following strategy.
(i) Since $\Sigma_{d d^{\dagger}}$ renormalizes
    in effect the energy of the level, we regard
    $\epsilon_d$ for the purpose of this section
    as implicitly containing its contribution,
    i.e., $\epsilon_d \to \epsilon_d +
    \Sigma_{d d^{\dagger}}$.
(ii) We neglect $\Sigma_{\psi \psi^{\dagger}}$
     altogether. Indeed, $\Sigma_{\psi \psi^{\dagger}}$
     corresponds to weak potential scattering
     $v_{\rm p}$, whose main effect is to slightly
     renormalize the conduction-electron density
     of states according to
     $\rho_0 \to \rho_0/[1 + (\rho_0 v_{\rm p})^2 ]$.
     We therefore expect the omission of
     $\Sigma_{\psi \psi^{\dagger}}$ to have only
     little effect on $\Gamma_{\rm eff}$.

With these simplifications, the calculation of
$\Gamma_{\rm eff}$ for nonzero $\epsilon_d$ closely
resembles its computation for $\epsilon_d = 0$.
Specifically, the Green function $G_{d\psi^{\dagger}}$
of Eq.~(\ref{equilibrium_G_dpsi}) acquires the modified
form
\begin{equation}
G_{d \psi^{\dagger}}(z) = \gamma^*_{\rm eff}
     \frac{ g_0(z) }
          { z - \epsilon_d - g_0(z)|\gamma_{\rm eff}|^2} ,
\end{equation}
which shifts the location of the poles in the summation
over the Matsubara frequencies in Eq.~(\ref{Matsubara}).
The self-consistency equation for $\Gamma_{\rm eff}$
remains given by Eq.~(\ref{SC-equilibrium}), however
$\Lambda(\Gamma_{\rm eff})$ is replaced by
\begin{equation}
\Lambda_{d}(\Gamma_{\rm eff}) =
   {\rm Re} \biggl \{
                      \frac{1}{x_{d}}
                      \frac{1}{1 + i\epsilon_d/D}
                      \left[
                               \psi(z_{+})
                               - \psi(z_{-})
                      \right]
            \biggr \} ,
\end{equation}
where
\begin{equation}
z_{\pm} = \frac{1}{2} +
          (1 \pm x_{d}) \frac{\beta (D + i\epsilon_d)}
                             {4\pi}
\end{equation}
and
\begin{equation}
x_{d} = \sqrt{ 1 - \frac{4}{1 + i\epsilon_d/D}
                   \frac{\Gamma_{\rm eff} + i\epsilon_d}
                        {D + i\epsilon_d} } .
\end{equation}
As in the previous sections, we exploit the largeness of
$D$ to expand in $\epsilon_d/D$, $\Gamma_{\rm eff}/D$,
and $T/D$. Keeping only the leading terms results in
\begin{equation}
\Lambda_{d}(\Gamma_{\rm eff}) \simeq
    {\rm Re} \left \{
                      \psi \left(
                                   \frac{\beta D}{2\pi}
                           \right)
                    - \psi \left(
                                   \frac{1}{2}
                                   + \beta
                                     \frac{\Gamma_{\rm eff}
                                           + i\epsilon_d}
                                          {2\pi}
                           \right)
             \right \} ,
\label{Lambda-finite-e}
\end{equation}
which generalizes Eq.~(\ref{Lambda-finite-T}) to
nonzero $\epsilon_d$.

Now the interplay between $\epsilon_d$, $\Gamma_{\rm eff}$,
and $T$ can now be read off from the argument of the
second digamma function in Eq.~(\ref{Lambda-finite-e}).
For $\beta|\Gamma_{\rm eff} +i\epsilon_d|\ll 1$
particle-hole symmetry breaking is irrelevant
and  Eq.~(\ref{SC-G-vs-T})
is recovered. For $\beta|\Gamma_{\rm eff} +i\epsilon_d|\gg 1$, the asymptotic 
expansion of $\psi(x)$ yields
\begin{equation}
\Lambda_{d}(\Gamma_{\rm eff}) \simeq
    \ln \left(
                \frac{D}
                     {\sqrt{\Gamma_{\rm eff}^2
                      + \epsilon_d^2}}
        \right) ,
\label{Lambda-as-log-ed}
\end{equation}
which generalizes Eq.~(\ref{Lambda-as-log})
to nonzero $\epsilon_d$ by replacing
\begin{eqnarray}
\Gamma_{\rm eff}&\to&
|\Gamma_{\rm eff} +i\epsilon_d|=\sqrt{\Gamma_{\rm eff}^2
                      + \epsilon_d^2} 
                      \, .
\end{eqnarray}
While for $\Gamma_{\rm eff} \gg |\epsilon_d|$, Eq.~(\ref{Lambda-as-log}) is approached, and for $|\epsilon_d| \gg \Gamma_{\rm eff}$, however,
Eq.~(\ref{Lambda-as-log-ed}) reduces to
\begin{equation}
\Lambda_{d}(\Gamma_{\rm eff}) \simeq
    \ln \left(
                \frac{D}
                     {|\epsilon_d|}
        \right)  \,
\end{equation}
and we obtain
\begin{equation}
\Gamma_{\rm eff} \simeq
       \frac{\Gamma_0}
            {\left[
                     1 - \rho_0 U \ln( D/|\epsilon_d| )
             \right]^2} ,
\end{equation}
or equivalently
\begin{equation}
\Gamma_{\rm eff}(\epsilon_d)
        \simeq \Gamma_0
               \left(
                       \frac{D}{|\epsilon_d|}
               \right)^{2 \rho_0 U}
\end{equation}
provided $\rho_0 U \ln( D/|\epsilon_d| ) \ll 1$.
$|\epsilon_d|$ serves in this case as the effective low-energy cutoff.

\subsection{Numerical Results}

We now turn to treat the general case, and present here the 
numerical solution of 
Eq.~(\ref{SC-equilibrium}) describing the general behavior of $\Gamma_{\rm eff}$ for different bare parameters of the model in our conserving approximation. 

\begin{figure}[tbf]
\centerline{
\includegraphics[width=85mm]{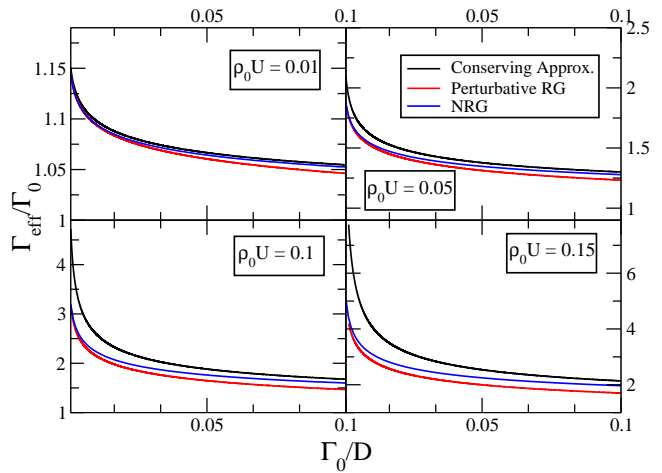}
}\vspace{0pt}
\caption{(Color online)
        The broadening of the level width $\Gamma_{\rm eff}/\Gamma_0$ as a function of the bare level width ${\Gamma_0}$ for different values of the coupling parameter $\rho_0 U$. For each value we present the broadening as calculated by solving the conserving approximation expression of Eq.~(\ref{SC-equilibrium}) (black), as calculated by applying the leading-order perturbative RG scaling of Eq.~(\ref{G_eff-RG-II}) (red), and using Wilson's NRG (blue).}
\label{Fig:zero_temp}
\end{figure}

In figure \ref{Fig:zero_temp} we compare the broadening of the
$d$-level at zero temperature obtained from the self-consistent solution
of Eq.~(\ref{SC-equilibrium}) and the analytical approximate solution in
Eq.~(\ref{G_eff-RG-II}), consistent with leading order perturbative RG. We augment these two sets of analytical data with results obtained using
Wilson's numerical renormalisation group (NRG) approach\cite{Wilson75,BullaCostiPruschke2008} which includes $\rho_0 U$ to all orders. In order to avoid discretization errors, we have extracted the renormalised parameters \cite{HewsonOguriMeyer2004} directly from the NRG fixed-point spectra of the IRLM \cite{JovchevAnders2013}. For small values of the coupling $\rho_0 U$, all approaches agree in the wide band limes. As $\rho_0 U$ increases, the conserving approximation differs quantitatively from the result predicted by the perturbative RG. While NRG and perturbative RG agree nicely for small $\Gamma_0/D$, i.\ e.\ in the wide band limit, significant deviations are observed for decreasing band width.
For increasing $\Gamma_0/D$, the leading perturbative RG underestimates
$\Gamma_{\rm eff}$ while the conserving approximation slightly  overestimates the renormalisation of the level broadening. Nevertheless, the NRG data seems to approach the results of the conserving approximation for increasing $\Gamma_0/D$, indicating that it also includes higher order contributions in $\rho_0 U$ due to the self-consistency condition. A significant increase of the broadening is observed even for $\rho_0 U = 0.1$. We note that for $\rho_0 U = 0.1$ and $\Gamma_0/D=0.1$, the conserving approximation differs only by less than 10\%
indicating that our approach describes well the physics in this regime
where the interaction plays an important role.

\begin{figure}[tbf]
\centerline{
\includegraphics[width=85mm]{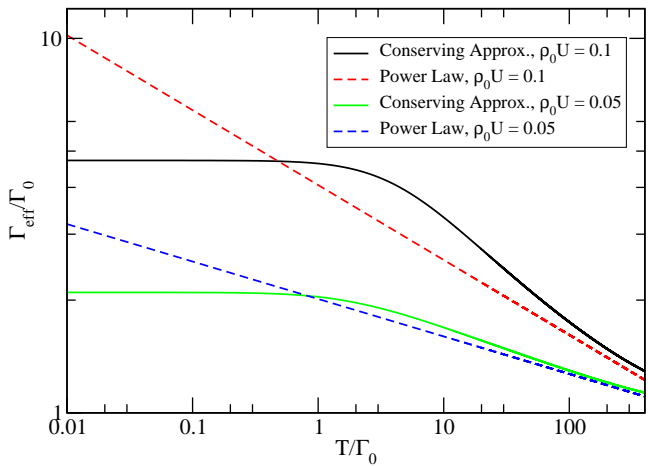}
}\vspace{0pt}
\caption{(Color online)
        The broadening of the level width $\Gamma_{\rm eff}/\Gamma_0$ as a function of the temperature, for different values of the coupling $\rho_0 U$. The continuous lines represent the numerical solutions of Eq.~(\ref{SC-equilibrium}) for finite temperature, and the dashed lines are the power-law behavior for high temperatures described by Eq.~(\ref{SC-G-vs-T}). Here $\Gamma_0/D = 10^{-3}$.}
\label{Fig:temp_depend}
\end{figure}

The temperature dependence of the level broadening on the temperature is plotted in Fig.~\ref{Fig:temp_depend}. For low temperatures with respect to $\Gamma_{\rm eff}$, the broadening is almost temperature independent. Once $T$ exceeds $\Gamma_{\rm eff}$, the graph converges to the power-law behavior predicted by Eq.~(\ref{SC-G-vs-T}): Our approach is qualitatively and quantitatively in agreement with the RG results which has been derived using an effective low-energy cutoff in the RG equation of ${\rm max}\{T,\Gamma_{\rm eff},|\epsilon_d|\}$.

\begin{figure}[tbf]
\centerline{
\includegraphics[width=85mm]{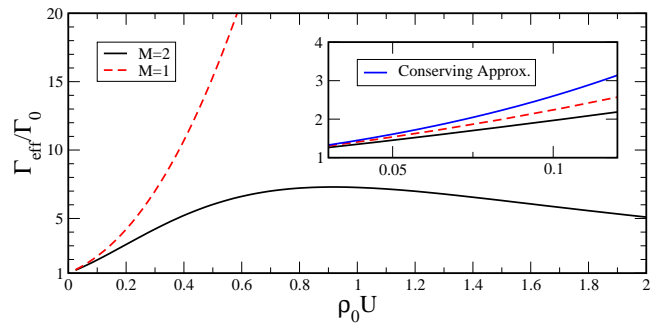}
}
\vspace{0pt}
\caption{
(Color online)
The ratio $\Gamma_{\rm eff}/\Gamma_0$  versus $g=\rho_0 U$ calculated with the NRG in equilibrium for two screening
channels ($M=2$) and $T\to 0$ for an band width $D/\Gamma_0=100$. Inset: the weak interacting range with the broadening ratio from the conserving approximation (blue).
\label{Fig:two-channel-gamma-effective}
}
\end{figure}

While $\Gamma_{\rm eff}$ monotonically increases with
increasing coupling constant $g=\rho_0 U$ for a single screening channel,
it reaches a maximum at the duality point\cite{SchillerAndrei2007-I}
 for $M=2$ after which $\Gamma_{\rm eff}$ declines again for larger
$\rho_0 U$.  This is illustrated by the equilibrium NRG data  
for the ratio  $\Gamma_{\rm eff}/\Gamma_0$
presented in Fig.~\ref{Fig:two-channel-gamma-effective}. As in Fig.\ \ref{Fig:zero_temp}, the NRG data has been obtained from the NRG fixed point spectra. \cite{HewsonOguriMeyer2004,JovchevAnders2013} The duality point is located approximately at $\rho_0 U=U/2D\approx 0.8$ within the NRG and differs slightly from the factor $2/\pi$ from the bosonisation treatment\cite{SchillerAndrei2007-I}
due to different cutoffs. While this point is 
independent of $D$, the absolute values of  the 
ratio $\Gamma_{\rm eff}/\Gamma_0$ is band width dependent as already demonstrated in Fig.~(\ref{Fig:zero_temp}).

While we have neglected the remaining two-particle interaction in our Green function approach, 
this interaction remains  present in the equilibrium NRG approach. 
Close to the Fermi-liquid fixed point perturbations can only contain irrelevant operators,
apart from one magical operator breaking particle-hole symmetry
as has been analytically worked out  in detail in Ref.\ (\onlinecite{KrishWilWilson80a}). 
The overall dimensionless strength\cite{KrishWilWilson80a} $\omega_2$ 
of the  leading order particle-particle interaction, scaling as $\Lambda^{-(N-1)/2}$ with the NRG iteration $N$, where $\Lambda>1$ is the NRG discretization parameter, measures the
degree of correlations and also enters the Wilson ratio.\cite{KrishWilWilson80a} 
We have extracted 
$\omega_2$ from the NRG level flow for the two-lead IRLM and found, that for $0\le \rho_0 U <0.2$,
$\omega_2$ is very small and, for $\rho_0 U=0.2$, corresponds to the residual particle-particle interaction strength found
in a symmetric single impurity Anderson model \cite{KrishWilWilson80a}  (SIAM) in the very weakly
correlated regime of $U/\Gamma_{\rm eff} \approx 0.1$. 
This justifies the neglect of the residual particle-particle interaction in our weak coupling approach to non-equilibrium as
presented here.

In contrary to $U>0$, where $\Gamma_{\rm eff}>\Gamma_0$, $\Gamma_{\rm eff}$ decreases
for negative $U$ and takes on the role of the Kondo temperature $T_K\propto \Gamma_{\rm eff}\ll \Gamma_0$.
In this regime, a completely different picture emerges:\cite{Schlottmann1978} the residual particle-particle 
interaction increases to large values as found in the  SIAM for $U/\Gamma_{\rm eff} \gg 1$
when approaching the quantum critical point.\cite{Schlottmann1978} This strongly correlated regime,
however, is not subject of investigation here.

\section{Nonequilibrium steady state}
\label{sec:noneq_steady_state}

\subsection{Conserving approximation at finite bias}

Now we extend the IRLM of Eq.~(\ref{H_general})
to two leads, i.\ e.\  $M=2$, each held at different
chemical potential and calculate the steady-state
current through the resonant level as function of
the bias voltage. For that purpose, we employ
the same conserving approximation as introduced
in Sec.~\ref{sec:conserving-approximation}
and calculate the renormalized  bias
dependent hybridization widths.

To simplify the calculation and to tune the system in the regime
of the strongest non-equilibrium effects we will focus on 
the symmetrical case where $\gamma_L=\gamma_R=\gamma$.

The symmetrized current operator from the left to the right
lead can be derived from the change of particle numbers between right and 
left lead~\cite{units}: 
\begin{eqnarray}
	\hat{I} &=& e\frac{i}{2}
	\left[\hat{N}_R-\hat{N}_L,\mathcal{H}\right] =
	\nonumber \\ &&
	-ie\frac{\gamma}{2}
	\left[\psi^{\dagger}_Rd-d^{\dagger}\psi_R-
	\psi^{\dagger}_Ld+d^{\dagger}\psi_L
	\right],
\end{eqnarray}
where $(-e)$ is the electrons charge, $\hat{N}_\alpha=\sum_k
c^{\dagger}_{\alpha\; k}c_{\alpha\; k}$ are the operators
for the number of electrons in each lead and
\begin{equation}
	\psi^{\dagger}_{\alpha} = 
	\frac{1}{\sqrt{N}}\sum_k c^{\dagger}_{\alpha\; k},
\end{equation}
is the local conduction electron in the lead $\alpha$ at the $d$-orbital. The steady-state current $I=\langle \hat{I} \rangle$ is then given by
\begin{equation}
	I = e\gamma{\rm Im}
	\left[G^{<}_{d\psi^{\dagger}_R}(t,t)-
	G^{<}_{d\psi^{\dagger}_L}(t,t)\right],
	\label{steady_state_current_real_time}
\end{equation}
and is related to the off-diagonal lesser Green function
\begin{equation}
	G^{<}_{d\psi^{\dagger}_\alpha}(t,t') =
	\langle \psi^{\dagger}_\alpha(t')d(t)\rangle.
	\label{eq:lesser-gf-49}
\end{equation}
In the steady-state, we can make use of the translational invariance
in time, i.e.\  $G^<(t,t')=G^<(t-t')$, and expand the current 
in the single frequency Fourier
representation of the equal time Green function
\begin{equation}
	I = e\gamma {\rm Im}\left\{
	\int\!\frac{d\epsilon}{2\pi}
	\left[G^{<}_{d\psi^{\dagger}_R}(\epsilon)-
	G^{<}_{d\psi^{\dagger}_L}(\epsilon)
	\right]
	\right\}.
	\label{steady_state_current}
\end{equation}

For the two-lead problem, it is useful to extend the $2\times 2$ matrices to $3\times 3$ matrices
\begin{equation}
\mat{G}^{\nu}(\epsilon) = \left[
               \begin{array}{ccc}
                     G^{\nu}_{\psi_L \psi_L^{\dagger}}(\epsilon) &
							G^{\nu}_{\psi_L \psi_R^{\dagger}}(\epsilon) &
                     G^{\nu}_{\psi_L d^{\dagger}}(\epsilon) \\ \\
                     G^{\nu}_{\psi_R \psi_L^{\dagger}}(\epsilon) &
							G^{\nu}_{\psi_R \psi_R^{\dagger}}(\epsilon) &
                     G^{\nu}_{\psi_R d^{\dagger}}(\epsilon) \\ \\
                     G^{\nu}_{d \psi_L^{\dagger}}(\epsilon) &
                     G^{\nu}_{d \psi_R^{\dagger}}(\epsilon) &
                     G^{\nu}_{d d^{\dagger}}(\epsilon)
               \end{array}
       \right] ,
	\label{G_3x3_defined}
\end{equation}
for the retarded ($\nu=r$), advanced ($\nu=a$), and the
lesser ($\nu=<$) Green functions.

The fully dressed retarded and advanced Green
function matrix is obtained for the formal solution of a Dyson equation
\begin{equation}
	\mat{G}^{r,a}(\epsilon) = \left[[\mat{G}^{r,a}_0(\epsilon)]^{-1} -
								\mat{\Sigma}(\epsilon) \right]^{-1},
\end{equation}
where $\mat{G}^{r,a}_0(\epsilon)$ is the non-interacting Green
function matrix and the components of the self-energies $\mat{\Sigma}(\epsilon)$
are derived from the generating functional.

Within the self-consistent
approximation these self-energies remain static and
independent of energy in the steady-state non-equilibrium case
\begin{eqnarray}
 \Sigma_{d d^{\dagger}} &=&
        U \langle :\!\psi_L^{\dagger} \psi_L\!: \rangle+
			U \langle :\!\psi_R^{\dagger} \psi_R\!: \rangle
			\\
 \Sigma_{\psi_{\alpha} \psi^{\dagger}_{\alpha}} &=& 
        U \langle d^{\dagger}d-\frac{1}{2}\rangle,
	\\
 \Sigma_{\psi_{\alpha} d^{\dagger}} &=& \Sigma_{d \psi_{\alpha}^{\dagger}}^*
     =\gamma - U\langle \psi_{\alpha}^{\dagger} d \rangle.
\end{eqnarray}
For symmetric couplings $\gamma_R=\gamma_L=\gamma$, and a structureless particle-hole symmetric density of states for both leads, we can focus on a symmetric voltage bias $\mu_L=-\mu_R=V/2$. The problem remains particle-hole symmetric for $\epsilon_d=0$, if one
interchanges the left and right leads in the process, {\it
i.e.} under the transformation $c_{L,R\; k} \to
c^{\dagger}_{R,L\; -k}$, $d\to -d^{\dagger}$. This
symmetry constrains the expectation values to
\begin{align}
	& \langle d^{\dagger}d \rangle = \frac{1}{2}
	\\
	&	\langle :\!\psi^{\dagger}_L\psi_L\!:\rangle +
	\langle :\!\psi^{\dagger}_R\psi_R\!:\rangle = 0,
\end{align}
and consequently
$\Sigma_{d d^{\dagger}}$ and 
$\Sigma_{\psi_{\alpha} \psi^{\dagger}_{\alpha}}$ vanish identically as in equilibrium. Finally, defining a lead dependent tunneling matrix element
\begin{equation}
	\gamma_{\rm eff}^{(\alpha)} =
	\gamma - U \langle \psi^{\dagger}_{\alpha}d\rangle,
\end{equation}
the retarded and advanced Green functions are given by
\begin{equation}
	\mat{G}^{r,a}(\epsilon) = \left[
               \begin{array}{ccc}
                     [g^{r,a}_0(\epsilon)]^{-1} &
							0 &
                     -\gamma^{(L)}_{\rm eff} \\ \\
                     0 &
							[g^{r,a}_0(\epsilon)]^{-1} &
                     -\gamma^{(R)}_{\rm eff} \\ \\
                     -(\gamma^{(L)}_{\rm eff})^* &
                     -(\gamma^{(R)}_{\rm eff})^* &
                     \epsilon\pm i\eta
               \end{array}
       \right]^{-1} ,
\end{equation}
where $g^{r,a}_0(\epsilon)$ denotes the bare retarded or
advanced Green function pertaining to $\psi_{\alpha}$,
defined in Eq.~(\ref{g_bare}).

In order to obtain closed analytical results, we again assume a Lorentzian density-of-states
\begin{equation}
	g^{r,a}_0(\epsilon) = 
			\pi\rho_0
		\frac{1}{\epsilon/D \pm i},
\end{equation}
and employing a wide band limit $D\gg \gamma$.

We employ the Langreth rules~\cite{Langreth76} to relate the lesser Green function matrix $\mat{G}^<(\epsilon)$ to the fully dressed advanced and retarded Green functions
\begin{equation}
	\mat{G}^{<}(\epsilon) = \mat{G}^r(\epsilon)[\mat{g}^{r}(\epsilon)]^{-1}
	\mat{g}^<(\epsilon)[\mat{g}^{a}(\epsilon)]^{-1}\mat{G}^{a}(\epsilon),
	\label{eq_Langreth}
\end{equation}
where $\mat{g}^\nu(\epsilon)$ are the bare Green functions
matrices, given by
\begin{equation}
	\mat{g}^{r,a}(\epsilon) = \left[
               \begin{array}{ccc}
                     g^{r,a}_0(\epsilon) &
							 &
                      \\ \\
                      &
							g^{r,a}_0(\epsilon) &
                      \\ \\
                      &
                      &
                     \left(\epsilon\pm i\eta\right)^{-1}
               \end{array}
       \right] ,
\end{equation}
and the unperturbed lesser Green function matrix given by
\begin{equation}
	\mat{g}^{<}(\epsilon) = 2\pi\left[
               \begin{array}{ccc}
                     f_L(\epsilon)\rho(\epsilon-\mu_L) &
							 &
                      \\ \\
                      &
							f_R(\epsilon)\rho(\epsilon-\mu_R) &
                      \\ \\
                      &
                      &
                     \frac{1}{2}\delta(\epsilon)
               \end{array}
       \right] ,
	\label{eq_bare_g_lesser}
\end{equation}
with $f_{\alpha}(\epsilon)=f(\epsilon-\mu_\alpha)$ the
Fermi-Dirac distribution.

As in equilibrium, the shift of the off-diagonal self-energy in the presence of the Coulomb repulsion is related to the fully dressed off-diagonal lesser Green function
\begin{eqnarray}
	\gamma^{(\alpha)}_{\rm eff} - \gamma &=& -U\int\!\frac{d\epsilon}{2\pi}
										G^{<}_{d\psi^{\dagger}_{\alpha}}
										(\epsilon) = \nonumber \\ &&
				-U\gamma_{\alpha\;\rm eff}
				\int\!\frac{d\epsilon}{2\pi}\left[
				G^r_{dd^{\dagger}}(\epsilon)
				g^<_{\alpha}(\epsilon)
				+
				G^<_{dd^{\dagger}}(\epsilon)
				g^a_0(\epsilon)
				\right], \nonumber \\ &&
	\label{eq_g_lesser_steadystate}
\end{eqnarray}
defining the self-consistency equation for $\gamma^{(\alpha)}_{\rm eff}$. The particle-hole symmetry of the Hamiltonian combined with the interchanging of
the left and right-leads, requires that
\begin{equation}
	\Sigma_{\psi_{L}d^{\dagger}} =
	t-U\langle \psi^{\dagger}_Ld\rangle =
	t-U\langle d^{\dagger}\psi_R\rangle =
	\Sigma_{\psi_Rd^{\dagger}}^*,
\end{equation}
which translates to
\begin{equation}
	\gamma^{(L)}_{\rm eff} = (\gamma^{R}_{\rm eff})^*.
\end{equation}
This relation renders the two equations determining
$\gamma^{(L)}_{\rm eff}$ and $\gamma^{(R)}_{\rm eff}$ to be complex
conjugate of one another. In the two-lead case, the combined hybridization width of the level is given by
\begin{equation}
	\bar{\Gamma}_{\rm eff} = 
	\pi\rho_0 \left(|\gamma^{(L)}_{\rm eff}|^2
	+|\gamma^{(R)}_{\rm eff}|^2
	\right),
\end{equation}
and $\bar{\Gamma}_0 = 2\pi\rho_0\gamma^2$ denotes the hybridization
width at $U=0$.

Carrying out the integral in Eq.~(\ref{eq_g_lesser_steadystate}) 
requires some lengthy
analytical calculations, which we shall skip here and present only
the end result
\begin{equation}
	\gamma^{(\alpha)}_{\rm eff} =
	\frac{\gamma}
	{1-\rho_0 U \Lambda_{\alpha}(\bar{\Gamma}_{\rm eff})}.
\end{equation}
and subsequently 
\begin{equation}
	\bar{\Gamma}_{\rm eff} =
	\frac{\bar{\Gamma}_0}{|1-\rho_0 U \Lambda_{\alpha}
	(\bar{\Gamma}_{\rm eff})|^2}.
	\label{gamma_steady_state}
\end{equation}
Here, the function $\Lambda_{\alpha}(\bar{\Gamma}_{\rm eff})$ depends
on the voltage difference and requires
$\Lambda_{R}(\bar{\Gamma}_{\rm eff})=
\Lambda_{L}(\bar{\Gamma}_{\rm eff})^*$. This 
function $\Lambda_{R}(\bar{\Gamma}_{\rm eff})$ is given
by the analytic expression
\begin{eqnarray}
	\Lambda_R(\bar{\Gamma}_{\rm eff}) &=&
	\frac{i}{2}
	{\rm Im}
	\left\{\frac{4}{x}\frac{\psi(z_+)}{3+x}
	-\frac{4}{x}\frac{\psi(z_-)}{3-x}
	+
	\frac{\psi(y)}{1+\bar{\Gamma}/2D}
	\right\}
	+ \nonumber \\ &&
	{\rm Re}
	\left\{\frac{\psi(z_+)-\psi(z_-)}{x}\right\},
	\label{eq_lambda_R}
\end{eqnarray}
where we have introduced the shorthand notations
\begin{eqnarray}
	4\pi z_{\pm} &=& 2\pi-i\beta V+(1\pm x)\beta D,
	\nonumber \\
	4\pi y &=& 2\pi-i\beta V + 2\beta D,
\end{eqnarray}
and defined $x=\sqrt{1-4\bar{\Gamma}_{\rm eff}/D}$.

We substitute these results in the expression of the steady-state current, i.\ e.\ Eq.~(\ref{steady_state_current}), and obtain one central result of our paper:
\begin{equation}
	I = e\frac{\gamma}{U}{\rm Im}\left\{
	\gamma^{(R)}_{\rm eff}-\gamma^{(L)}_{\rm eff}
	\right\} = 
	2G_0\bar{\Gamma}_{\rm eff}
	{\rm Im}\{\Lambda_R(\bar{\Gamma}_{\rm eff})\},
	\label{steady_state_current_by_gamma}
\end{equation}
where $G_0=e/h$ is the fundamental quantum conductance. Note that in equilibrium, $\gamma^{(\alpha)}_{\rm eff}$ is real, and the current vanishes. A finite bias breaks time-reversal symmetry and $\gamma^{(\alpha)}_{\rm eff}$ becomes complex. Consequently a current can flow.

\subsubsection{Zero temperature limit}

Generally, Eq.~(\ref{gamma_steady_state}) has to be
solved numerically, and then the current is calculated
directly by plugging $\bar{\Gamma}_{\rm eff}$ into the
expression in Eq.~(\ref{steady_state_current_by_gamma}).
However, we find it useful to derive some analytical results for the zero temperature limit, $T\to 0$, first.

Exploiting the fact that $D \gg \bar{\Gamma}_{\rm eff}$, we expand $x\simeq 1 - 2\bar{\Gamma}_{\rm eff}/D$ in Eq.~(\ref{eq_lambda_R}) and arrive at the approximated expression
\begin{eqnarray}
	\Lambda_{R}(\bar{\Gamma}_{\rm eff}) &\simeq&
	\psi\left(\frac{1}{2}-i\frac{\beta V}{4\pi}+\frac{\beta D}{2\pi} \right) -
	\nonumber \\ &&
	\psi\left(\frac{1}{2}-i\frac{\beta V}{4\pi}+
	\frac{\beta\bar{\Gamma}_{\rm eff}}{2\pi} \right).
	\label{noneq_finite_T}
\end{eqnarray}
We use the expansion of the 
digamma function in Eq.~(\ref{eq:digamma-expansion})
for $T\to 0$ and are left with the expression
\begin{equation}
	\Lambda_R(\bar{\Gamma}_{\rm eff}) 
	\simeq \ln\left(
		\frac{D-iV/2}{\bar{\Gamma}_{\rm eff}-iV/2}
	\right),
	\label{Lambda_zero_temp}
\end{equation}
from which we can derive the approximated current using Eq.~(\ref{steady_state_current_by_gamma})
\begin{equation}
	I \simeq 2G_0\bar{\Gamma}_{\rm eff}
	\tan^{-1}\left[\frac{V}{2\bar{\Gamma}_{\rm eff}}
	\frac{1}{1+V^2/(4D\bar{\Gamma}_{\rm eff})}
	\right].
\end{equation}

For very low voltages $V \ll \bar{\Gamma}_{\rm eff} \ll D$, this expression reduces to a linear form
\begin{equation}
	I = G_0 V,
	\label{IV_linear}
\end{equation}
which is independent of the level width, reproducing the perfect transmission with conductance $G_0$ of a symmetric junction and ballistic transport for $V\to 0$. This result is not surprising since the equilibrium fixed point is a Fermi liquid where $U$ is dressing $\bar{\Gamma}_0$ to
$\bar{\Gamma}_{\rm eff}$ and determining the energy scale.

For increasing values of the voltage, $V \sim \bar{\Gamma}_{\rm eff} \ll D$, the current is approximated by
\begin{equation}
	\frac{I}{2\bar{\Gamma}_{\rm eff}} \simeq G_0 \tan^{-1}
	\left( \frac{V}{2\bar{\Gamma}_{\rm eff}} \right),
	\label{IV_universal_regime}
\end{equation} 
where we have written it in a universal form, characterized by a single energy scale $\bar{\Gamma}_{\rm eff}$.

When $V$ exceeds the effective level width, we substituted Eq.~(\ref{Lambda_zero_temp}) after neglecting $\bar{\Gamma}_{\rm eff}$ in the argument of the logarithm into Eq.~(\ref{gamma_steady_state}) and derive
\begin{equation}
	\bar{\Gamma}_{\rm eff} \simeq 
	\frac{\bar{\Gamma}_0}{\left|1-\rho_0 U
	\ln(2iD/V+1)\right|^2}.
\end{equation}
In the weak coupling limit $\rho_0 U \ll 1$, we employ the same approximation of the denominator as in Sec.~\ref{sec:zero_temp_eq} and obtain
\begin{equation}
	\bar{\Gamma}_{\rm eff} \simeq \bar{\Gamma}_0
	\left(\frac{V}{\sqrt{4D^2+V^2}}\right)^{-2\rho_0 U},
	\label{eq:gamma-eff-vs-bias}
\end{equation}
so that the current is given by
\begin{equation}
	I \simeq  
	G_0\bar{\Gamma}_{\rm eff}\left[
	\pi-2\tan^{-1}\left(\frac{V}{2D}\right)
	\right],
	\label{IV_high_V}
\end{equation}
in this limit.

Within the regime $\bar{\Gamma}_{\rm eff} \ll V\ll D$, the voltage $V$ serves as the low-energy cutoff. The effective width of the level has a power-law dependence on the voltage with an exponent $(-2\rho_0 U)$: It plays a similar role as the temperature in equilibrium. When $V$ approaches the band width $D$, the width of the level experience almost no renormalization, and remains at its bare value $\bar{\Gamma}_0$.

For $V \gg \bar{\Gamma}_{\rm eff}$, the current decreases with increasing bias, as the magnitude of both the effective level width and the imaginary part of $\gamma^{\alpha}_{\rm eff}$, described by the term in the parenthesis in Eq.~(\ref{IV_high_V}), decreases. Consequently,  this regime is characterized by a negative differential conductance,\cite{BoulatSaleurSchmitteckert2008,BordaZawa2010,RT-RG-IRLM-2,RT-RG-IRLM}
another central result of our paper.

The negative differential conductance has been reported in the literature using sophisticated state of the art numerical approaches\cite{BoulatSaleurSchmitteckert2008}
perturbative RG \cite{BordaZawa2010} or functional RG methods.\cite{RT-RG-IRLM-2,RT-RG-IRLM} Within our approach, it is to be understood as a twofold effect -- the decrease in the effective width of the level, caused by the fact that for high voltages (with respect to the width $\bar{\Gamma}_{\rm eff}$) the voltage serves as the low-energy cut-off of the renormalization process, and a decrease in the overlap between the bandwidths of the two leads, which is manifested by a decrease in the imaginary part of $\gamma^{(\alpha)}_{\rm eff}$. This latter effect is present even in the non-interacting case $U=0$, and is governed not by the size of $V$ with respect to the level width but with the size of $V$ with respect to the bandwidth $D$. This latter effect is not universal, but is both cutoff dependent and setup dependent.~\cite{Branschadel2010Ann} As such, its role is expected to be less significant within the physical regime, where we keep the electronic bandwidth as the largest energy scale of the system.

It is worth noting that as long as $T,V\ll \bar{\Gamma}_{\rm eff}$, the bandwidth $D$ does not play any role besides determining $\bar{\Gamma}_{\rm eff}$. The current is a universal function which scales with the effective width  $I/\bar{\Gamma}_{\rm eff} = f(V/\bar{\Gamma}_{\rm eff}, T/\bar{\Gamma}_{\rm eff})$. This ceases to be the case outside this parameters regime, and the values of $V$ and $T$ with respect to $D$ become important.

\subsubsection{Finite Temperature}
The previous discussion for $T=0$ can be readly extended to finite temperatures as long as $T\ll\bar{\Gamma}_{\rm eff}$. For temperatures outside this regime, as in the equilibrium case, the digamma function in Eq.~(\ref{noneq_finite_T}) cannot be reduced to log functions. The only additional complexity compare to the equilibrium is the finite voltage.

For simplicity, we restrict the discussion to the case where $T$ is much smaller than the bandwidth, allowing us to approximate the first digamma function in Eq.~(\ref{noneq_finite_T}) by a log function. The second digamma function will approach the constant value of $\psi(1/2)$ as we increase $T$ to be larger than both $V$ and $\bar{\Gamma}_{\rm eff}$. As such, when $T \gg  \max\{V,\bar{\Gamma}_{\rm eff}\}$, it serves as the low-energy cutoff and we can approximate the value of the effective width in a similar manner to the equilibrium case described in Eq.~(\ref{SC-G-vs-T}) for $\Gamma_{\rm eff}$. Increasing $T$ then reduces the magnitude of $\bar{\Gamma}_{\rm eff}$, which reduces the current through the level. 

We define the backscattered current as $I_{BS} = G_0 V - I$. Expanding it in the low voltage and low temperature regime, we get
\begin{equation}
	I_{BS} \simeq G_0 V
				\left(\frac{1}{12}\frac{V^2}{\bar{\Gamma}_{\rm eff}^2}
				+\pi\frac{T}{\bar{\Gamma}_{\rm eff}}+
				\frac{\pi}{4}\frac{TV^2}{\bar{\Gamma}_{\rm eff}^3}
				\right),
	\label{eq_I_BS}
\end{equation}
where we have considered the linear power in $T/\bar{\Gamma}_{\rm eff}$ and cubic power in $V/\bar{\Gamma}_{\rm eff}$, and employed the wide-band limit. This gives the leading temperature dependence of the backscattered current as well. The leading term, which is proportional to $(V/\bar{\Gamma}_{\rm eff})^3$, is consistent with Fermi liquid theory.~\cite{Freton2014} The interaction only plays a role in setting the energy scale $\bar{\Gamma}_{\rm eff}$.

\subsubsection{Numerical results}
To evaluate the steady-state current between the leads in the most general case, we solved Eq.~(\ref{gamma_steady_state}) numerically, and then plugged the result into Eq.~(\ref{steady_state_current_by_gamma}). Note that the finite bias voltage enters this self-consistency condition 
via Eq.~(\ref{eq_lambda_R}).

\begin{figure}[tbf]
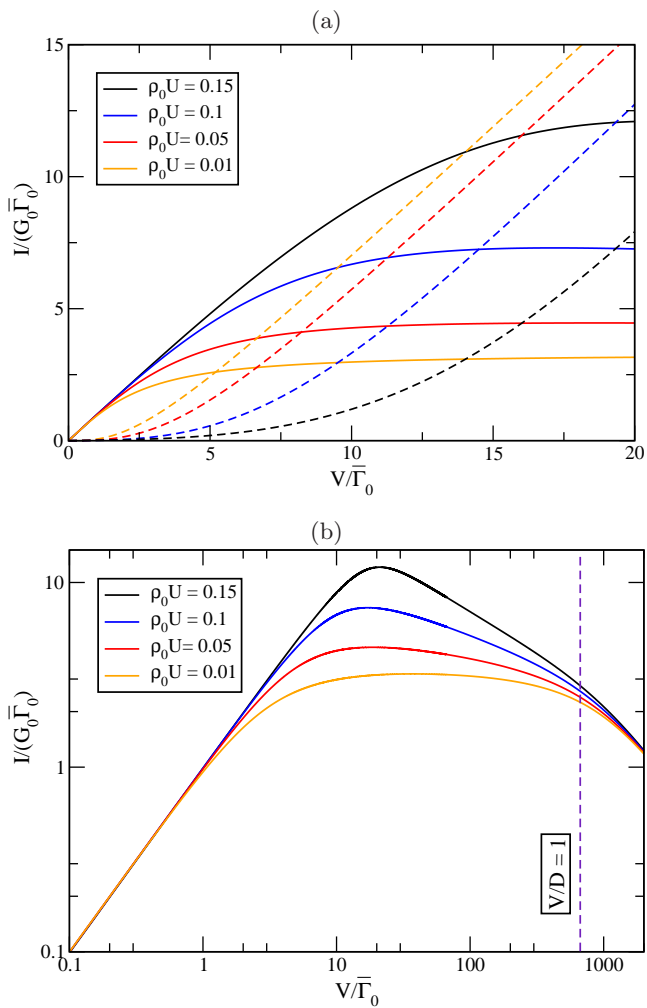


\begin{center}
(a) \includegraphics[clip,width=85mm]{IV_low_V.eps}

\vspace*{2mm}
(b) \includegraphics[clip,width=85mm]{IV_full_log.eps}

\end{center}

\caption{(Color online) 
(a) The  zero-temperature current between the leads as a function of the voltage (continuous lines), 
at zero temperature and in the low and intermediate voltage regime $V\sim \bar{\Gamma}_{\rm eff}$, for different values of the coupling $\rho_0 U$, and the backscattered current $I_{BS}$, in dashed lines. 
(b)
The same data as in (a) but for a larger range of $V$ and on an log-log scale.
The dashed perpendicular line represent the point where the voltage bias between the leads is equal to the lead-electrons bandwidth, $V=D$. Here $\bar{\Gamma}_0/D=1.5\cdot 10^{-3}$.
}
\label{Fig:IV_V}
\end{figure}

We start with the zero-temperature results. In figure \ref{Fig:IV_V}
we have plotted the zero-temperature current as a function of the voltage between the leads $V=\mu_R-\mu_L$ for the symmetrical case $\mu_L = -\mu_R$, $T=0$, and at resonance $\epsilon_d=0$, for different values of the interaction strength $\rho_0U$. In figure (\ref{Fig:IV_V})(a) the low-voltage behavior is presented, and the cross-over from the linear regime to the non-linear regime, both described by Eq.~(\ref{IV_universal_regime}), is evident. The cross-over occurs at different voltages depending on the coupling $\rho_0 U$: the cross-over scale is related to  $\bar{\Gamma}_{\rm eff}$ which is increasing with $U$. In this figure we have also plotted the backscattered current $I_{BS}$, and the slow cubic rise at low $V$, predicted by Eq.~(\ref{eq_I_BS}), is evident.

In figure~(\ref{Fig:IV_V})(b) the same data  as in Fig.~(\ref{Fig:IV_V})(a)  is presented, but for a larger range of voltages on a log-log scale. 
Here the negative-differential conductance at high-voltages as predicted
by Eq.~(\ref{IV_high_V}) is clearly visible. We have extended the bias to $D\ll V$: In that regime, seen at the far-right-side of the graph, 
all currents for the difference couplings converge to the same function  governed by the unrenormalized $\bar{\Gamma}_{0}$.
 
Figure~\ref{Fig:IV_V}(b) summarises one of the key findings of this paper: the leading order conserving approximation is sufficient to describe the negative differential conductance seen in much more sophisticated numerical approaches such as the TD-DMRG. \cite{BoulatSaleurSchmitteckert2008}.
For $\bar\Gamma_0 \ll V < D$, the current decays with a power law $V^{-2\rho_0 U}$ determined by the renormalization of $\Gamma_{\rm eff}$ and also consistent with a functional renormalization group approach\cite{RT-RG-IRLM}: the larger $U$ the larger the exponent, the faster the decay for increasing voltage. For large voltages, the current is governed by approach to unrenormalized charge fluctuation scale, and all current curves collapse.

\begin{figure}[tbf]
\centerline{
\includegraphics[width=85mm]{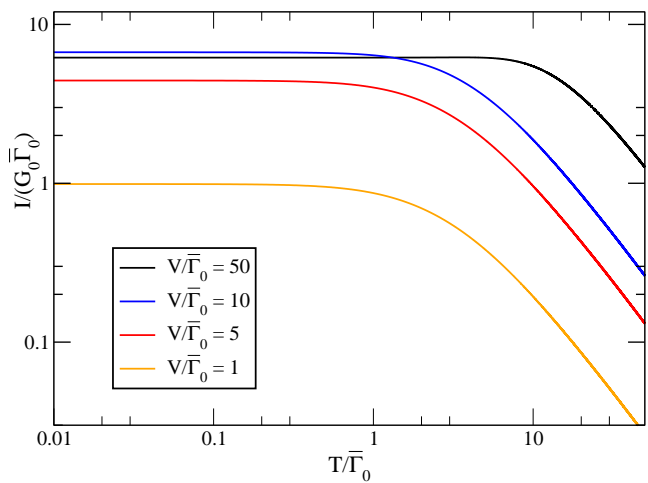}
}\vspace{0pt}
\caption{(Color online) The current between the leads as a function of the temperature, for different values of voltage bias $V$, on a log-log scale. Here $\bar{\Gamma}_0/D=1.5\cdot 10^{-3}$ and $\rho_0 U = 0.1$.}
\label{Fig:IT_log}
\end{figure}

Now we proceed to finite temperature. In figure (\ref{Fig:IT_log}) the temperature dependence of the current is plotted, for a single value of the coupling $\rho_0 U = 0.1$ and at different fixed voltages. The current remains temperature independent as long as $T\ll \bar{\Gamma}_{\rm eff}$.
Once the temperature exceeds the maximum of both $V$ and $\bar{\Gamma}_{\rm eff}$ a power-law decline of the current is observed.
This become particularly evident by comparing the lines pertaining to $V/\bar{\Gamma}_0 = 50$ and $V/\bar{\Gamma}_0 = 10$, which at low temperatures display similar values of the current (due to the negative differential conductance at hight voltages), but the latter starts decreasing, as we increase the temperature, much sooner than the former, which is more resilient due to the higher voltage.

\subsection{Shot Noise}

Shot noise measurements provide a direct indicator for correlation effects governing
quantum transport. The ratio between the shot noise $S_0$ and the current of the backscattered
particles $I_b$
\begin{eqnarray}
e^* &=& \frac{S_0}{2I_b}
\end{eqnarray}
has been used to define an effective charge of the quasiparticle responsible for  
the transport processes.  The most prominent examples are the fractional charge $e^*=e/3$
in the fractional quantum Hall regime \cite{FractionalCharge1997}, 
as well as the detection of the Cooper-pair
charge $e^*=2e$ in normal metal-superconductor junctions \cite{SuperconductorCharge1997}.
In the context of transport through a quantum dot in the strong coupling limit of the Kondo model,
characterised by a Wilson ratio\cite{Wilson75} $R=2$, Sela and collaborators have  reported \cite{SelanShotNoise2006} a fractional shot noise with $e^*/e= 5/3$ for $T,V\to 0$
which reduced to $e^*/e=1$ in the weak coupling limit, defined by $R\to 1$.

The zero-frequency shot noise at finite bias is defined by
\begin{equation}
	S_0 =
	\frac{1}{2}\int\!dt \langle\{ \delta\hat{I}(t)\delta\hat{I}(0)\}\rangle,
\end{equation}
where $\delta\hat{I} = \hat{I} - \langle\hat{I}\rangle$. 
We restrict ourselves to $T\to 0$, where $S_0$  reflects the quantum nature of the conductance
in the the absence of any thermal noise.

In order to calculate the shot noise, we need to supplement the lesser Green function stated in Eq.\ \eqref{eq:lesser-gf-49} by the corresponding
greater Green function
\begin{equation}
	G^>_{AB^{\dagger}}(t,t') = \langle B^{\dagger}(t')A(t) \rangle .
\end{equation}
Its Fourier transform with respect to the time difference $t-t'$, $G^>_{AB^{\dagger}}(\omega)$, 
can be calculated using Langreth's rules \cite{Langreth76}
\begin{equation}
	\mat{G}^{>}(\epsilon) = \mat{G}^r(\epsilon)[\mat{g}^{r}(\epsilon)]^{-1}
	\mat{g}^>(\epsilon)[\mat{g}^{a}(\epsilon)]^{-1}\mat{G}^{a}(\epsilon),
\end{equation}
where the bare greater Green functions given by
\begin{equation}
	\mat{g}^{>}(\epsilon) = 2\pi\left[
               \begin{array}{ccc}
                     \bar{f}_L(\epsilon)\rho(\epsilon-\mu_L) &
							 &
                      \\ \\
                      &
							\bar{f}_R(\epsilon)\rho(\epsilon-\mu_R) &
                      \\ \\
                      &
                      &
                     \frac{1}{2}\delta(\epsilon)
               \end{array}
       \right] ,
\end{equation}
with $\bar{f}_\alpha(\epsilon) = 1-f(\epsilon-\mu_{\alpha})$, and at zero temperature $\bar{f}_\alpha(\epsilon) = \theta(\epsilon-\mu_\alpha)$.

Within our conserving approximation, we apply Wick's theorem
\begin{eqnarray}
	\langle A^{\dagger}(t)B(t)C^{\dagger}(0)D(0) \rangle -
	\langle A^{\dagger}(t)B(t) \rangle \langle C^{\dagger}(0)D(0) \rangle
	&& \nonumber \\ \nonumber \\ \; \;  \;
		= G^<_{DA^{\dagger}}(t,0)G^>_{BC^{\dagger}}(0,t), &&
\end{eqnarray}
and only include the resummation of the single-particle terms by replacing the bare Green functions with
the fully-dressed propagators.
We also note that
\begin{equation}
	\int\!dt G^<_{DA^{\dagger}}(t,0)G^>_{BC^{\dagger}}(0,t) =
	\int\!\frac{d\omega}{2\pi}
	G^<_{DA^{\dagger}}(\omega)G^>_{BC^{\dagger}}(\omega).
\end{equation}

Additional diagrammatic corrections  would be included
into the irreducible two-particle vertex of a non-equilibrium particle-hole Bethe-Salpheter 
equation and are neglected here. Such terms would lead to additional backscattering 
contributions, modelled by a $\beta$-factor in Ref.~(\onlinecite{SelanShotNoise2006}).

Employing the wide-band limit $D\gg\bar{\Gamma}_{\rm eff},V$, and carrying out a rather lengthy calculation, the zero-frequency shot noise at zero temperature is given by
\begin{equation}
	S_0 
	= eG_0\frac{\bar{\Gamma}_{\rm eff}}{2}
	\left[2\tan^{-1}\left(\frac{V}{2\bar{\Gamma}_{\rm eff}}\right)
	-\frac{\bar{\Gamma}_{\rm eff}V}{\bar{\Gamma}_{\rm eff}^2+(V/2)^2}
	\right].
	\label{eq_shot_noise}
\end{equation}
This form is consistent with the general picture of the noise in a noninteracting setup with an level width defined by $\bar{\Gamma}_{\rm eff}$, where the transmission coefficient is $T(\epsilon) = \bar{\Gamma}_{\rm eff}^2/(\epsilon^2+\bar{\Gamma}_{\rm eff}^2)$. It reproduces the correct result in the noninteracting limit, where $\bar{\Gamma}_{\rm eff}=\bar{\Gamma}_0$.~\cite{Golub2007} At the range $V\ll \bar{\Gamma}_{\rm eff}$ the ratio between the noise and the backscattered current $I_{BS}$ is given by
\begin{equation}
e^* =  \frac{S_0}{2 I_{BS}} = e
\end{equation}
and therefore, effective charge remains unaltered from the bare charge.

Recently, the voltage dependent shot noise  was calculated exactly
for the IRLM using the Bethe {\it Ansatz} at the self-dual point 
$\rho_0 U = \pi/2$:
the ratio between the noise and the backscattered current in the low voltage regime yields
an effective charge of $e^* = 2e$, while for large bias voltages $e^* = e/2$ has been reported \cite{Branschadel2010,Carr2011}.
The enhancement of $e^*/e>1$ reflects the inclusion of two-particle scattering processes in the current and noise
calculation \cite{SelanShotNoise2006}. Such corrections have been neglected within our calculation. We also note that this range of strong interaction $\rho_0 U \sim 1$ lies well outside the range of validity of our approximation.
The value and voltage dependence of the shot noise ratio at the duality point indicates that the quasiparticle involving the transport have strongly modified properties compared to the weak coupling limit investigated here in this paper.

\section{Quench Dynamics}
We finally turn to consider the quench dynamics in the system under investigation. In a quench setup, the system is initially prepared in some equilibrium state (or steady-state), propagates with respect to a different Hamiltonian starting at some time $t_0$. This is modelled by an abrupt change of one or several of its parameters. In the general case, the system will be driven out of equilibrium and after some transitional period will relax to a new equilibrium or to a steady-state (though there are setups in which such systems do not reach even steady-state). In this section we will calculate the response of our system to different quenches, following the real-time dynamics as it approaches the steady-state or equilibrium state that has been described in the previous sections.

Before turning to address specific setups we present here a general discussion of our method, which in the literature is known as the time-dependent Hartree-Fock~\cite{Leeuwen2009}. As we are interested in following the real-time dynamics of physical observables, our goal is to calculate the expectation values of the type 
\begin{equation}
	G^<_{AB^{\dagger}}(t,t') = \langle B^{\dagger}(t') A(t) \rangle 
\end{equation}
at equal times $t=t'$, where $A$ and $B$ are fermionic operators pertaining to the degrees of freedom of the system, and we related it to the lesser Green function. In contrast to equilibrium or to nonequilibrium steady-state, the correlation functions following a quench are functions of two times, and not only of the time difference, not allowing a solution based on Fourier transforming to the energy domain. 

Similar to what was done in Eq.~(\ref{G_3x3_defined}) we define the Green function in matrix form $\mat{G}^{\nu}(t,t')$ for the retarded $(\nu=r)$, advanced $(\nu = a)$ and lesser $(\nu = <)$ functions. For the single-lead setup they will be $2\times 2$ matrices
\begin{equation}
\mat{G}^{\nu}(t,t') = \left[
               \begin{array}{cc}
                     G^{\nu}_{\psi \psi^{\dagger}}(t,t') &
                     G^{\nu}_{\psi d^{\dagger}}(t,t') \\ \\
                     G^{\nu}_{d \psi^{\dagger}}(t,t') &
                     G^{\nu}_{d d^{\dagger}}(t,t')
               \end{array}
       \right] ,
	\label{two_time_G_single}
\end{equation}
and in the case of a two-lead setup they will be $3\times 3$ matrices
\begin{equation}
\mat{G}^{\nu}(t,t') = \left[
               \begin{array}{ccc}
                     G^{\nu}_{\psi_L \psi_L^{\dagger}}(t,t') &
							G^{\nu}_{\psi_L \psi_R^{\dagger}}(t,t') &
                     G^{\nu}_{\psi_L d^{\dagger}}(t,t') \\ \\
                     G^{\nu}_{\psi_R \psi_L^{\dagger}}(t,t') &
							G^{\nu}_{\psi_R \psi_R^{\dagger}}(t,t') &
                     G^{\nu}_{\psi_R d^{\dagger}}(t,t') \\ \\
                     G^{\nu}_{d \psi_L^{\dagger}}(t,t') &
                     G^{\nu}_{d \psi_R^{\dagger}}(t,t') &
                     G^{\nu}_{d d^{\dagger}}(t,t')
               \end{array}
       \right].
	\label{two_time_G_two}
\end{equation}
Expanding the Green functions using regular perturbation series we can write
\begin{eqnarray}
	\mat{G}^{\nu}(t,t') &=& \int_{-\infty}^{\infty}d\tau_1 d\tau_2
						\left[\mat{G}(t,\tau_2)
					\mat{\Sigma}(\tau_2,\tau_1)\mat{g}_0(\tau_1,t')
						\right]^{\nu} \nonumber \\ &&
						+ \mat{g}^\nu_0(t,t'),
\end{eqnarray}
where $\mat{g}^{\nu}_0(t,t')$ is the bare Green function matrix. Next we exploit the fact that within our approximation, the self energies $\mat{\Sigma}(t,t')$ are instantaneous in time, leading to the form
\begin{equation}
	\mat{\Sigma}(t,t') = \mat{\Sigma}(t)\delta(t-t'),
\end{equation}
and rely on Langreth theorem~\cite{Langreth76} to expand explicitly the equations for the lesser Green functions
\begin{eqnarray}
	\mat{G}^<(t,t') &=& \int_{-\infty}^{\infty}d\tau 
	\big[\mat{G}^<(t,\tau)\mat{\Sigma}(\tau)\mat{g}^a_0(\tau,t') + 
	\nonumber \\ &&
	\mat{G}^r(t,\tau)\mat{\Sigma}(\tau)\mat{g}^<_0(\tau,t')
	\big] + \mat{g}^<_0(t,t'),
	\label{quench_eqs_G_<}
\end{eqnarray}
and the retarded Green functions
\begin{eqnarray}
	\mat{G}^r(t,t') &=& 
	\int_{-\infty}^{\infty}d\tau 
	\mat{G}^r(t,\tau)\mat{\Sigma}(\tau)\mat{g}^r_0(\tau,t') +
	\nonumber \\ &&
	\mat{g}^r_0(t,t').
	\label{quench_eqs_G_r}
\end{eqnarray}
The self energy matrix at time $\tau$, for the single-lead setup, is given by
\begin{equation}
	\mat{\Sigma}(\tau) = \left(
	\begin{array}{cc}
	U G^<_{dd^{\dagger}}(\tau,\tau) &
	\gamma - U G^<_{\psi d^{\dagger}}(\tau,\tau) \\ \\
	\gamma - U G^<_{d \psi^{\dagger}}(\tau,\tau) &
	U G^<_{\psi\psi^{\dagger}}(\tau,\tau)
	\end{array}
	\right),
	\label{self-energy-real-time-1lead}
\end{equation}
\begin{widetext}
and for the two-lead setup is given by
\begin{equation}
	\mat{\Sigma}(\tau) =
	\left[
	\begin{array}{ccc}
	U G^<_{dd^{\dagger}}(\tau,\tau) & 0 &
	\gamma - U G^<_{\psi_L d^{\dagger}}(\tau,\tau) \\ \\
	0 & U G^<_{dd^{\dagger}}(\tau,\tau) &
	\gamma - U G^<_{\psi_R d^{\dagger}}(\tau,\tau) \\ \\
	\gamma - U G^<_{d \psi_L^{\dagger}}(\tau,\tau) &
	\gamma - U G^<_{d \psi_R^{\dagger}}(\tau,\tau) &
	U \left(G^<_{\psi_L\psi_L^{\dagger}}(\tau,\tau)+
	G^<_{\psi_R\psi_R^{\dagger}}(\tau,\tau)\right)
	\end{array}
	\right].
	\label{self-energy-real-time-2leads}
\end{equation}
\end{widetext}

We are at a position to lay out the strategy for numerically solving the set of integrals equations in Eqs.~(\ref{quench_eqs_G_<}-\ref{quench_eqs_G_r}). 
All the bare Green functions, and also $\mat{\Sigma}(\tau)$ 
are known prior to the quench, i.e. at $\tau < 0$. Causality, encoded in the $\theta$ functions of the retarded and advanced Green functions, cut off the time arguments in the integrals in Eqs.~(\ref{quench_eqs_G_<}-\ref{quench_eqs_G_r}) in such a way that for $\mat{G}^{\nu}(t,t')$ with $t'\leq t$, only the self-energy at time $\tau\leq t$ enters the equations: Only the past enters the equations.

We define a discrete time step $\Delta t$, and assuming that we know $\mat{\Sigma}(\tau)$ for all $\tau \leq t-\Delta t$, we fix $t$ as a parameter. Equations~(\ref{quench_eqs_G_<}-\ref{quench_eqs_G_r}) are 
then self-consistent and solved numerically for $\mat{G}^<(t,t')$ and $\mat{G}^r(t,t')$ at the range $t'\leq t$. From this solution we calculate the next self-energy value $\mat{\Sigma}(t)$, setting the ground for repeating the process, this time solving $\mat{G}^<(t+\Delta t, t')$ and $\mat{G}^r(t+\Delta t, t')$ for $t' \leq t+\Delta t$. Starting with $\mat{G}(\Delta t, t')$, we iterate this process step-by-step until at long times we converge to the steady-state solution of Eqs.~(\ref{quench_eqs_G_<}-\ref{quench_eqs_G_r}) described in the previous sections, where all correlation functions are only functions of the time-difference. After this technical digression,
we turn to consider different specific quenches applied to the model and present the results.

\subsection{Connecting the level to a single-lead}
\label{sec:quench1-lead}
Let us consider a system composed of a  level initially decoupled from a single lead for times $t<0$. At $t=0$, they are connected  by turning on the hopping term in the Hamiltonian of Eq.~(\ref{H_general}). In this setup we will follow the time evolution of the effective width of the level $\Gamma_{\rm eff}(t)$ until it reaches its equilibrium value described ib Sec.~(\ref{sec:termal_eq}).

The system at times $t\leq 0$ is at thermal equilibrium with resepct to the disconnected Hamiltonian
\begin{eqnarray}
	\mathcal{H}_0 &=& \sum_k \epsilon_k c^{\dagger}_k c_k +
						\epsilon_d d^{\dagger}d 
	+ \nonumber \\ &&
	\frac{U}{N}
	\left(d^{\dagger}d-\frac{1}{2}\right)
	\sum_{k,q}
	:\!c^{\dagger}_{k}c_{q}\!:,
	\label{H_0_nohopping}
\end{eqnarray}
and the dynamics of the degrees of freedom is fully described by the bare Green functions $g^\nu_{0\;\psi\psi^{\dagger}}(t-t')$ and $g^\nu_{0\; dd^{\dagger}}(t-t')$ which are functions only of the time difference $\tau=t-t'$. As in section (\ref{sec:termal_eq}), we shall focus on the particle-hole symmetric case where the level is held at resonance $\epsilon_d=0$, and the density of states of the lead is symmetric with half-width $D$. Note that the Coulomb repulsion $U$ has been
absorbed into the definition of the bare parameters, as discussed in
connection with the Hartree equations (\ref{eq:sigma-dd}-\ref{eq:sigma-psi-d}). We will focus on the zero-temperature limit $T\to 0$, and will extend our theory
to finite temperature later.

Under these conditions, the bare Green functions of the dot degrees of freedom are given by
\begin{eqnarray}
	g^r_{0\; dd^{\dagger}}(\tau) &=& -i\theta(\tau), \nonumber \\
	g^<_{0\; dd^{\dagger}}(\tau) &=& \frac{1}{2},
	\label{bare_g_dot}
\end{eqnarray}
and the bare Green functions pertaining to the electronic lead degree of freedom $\psi$, for a Lorentzian density of states and at zero temperature, are
\begin{eqnarray}
	g^r_{0\; \psi\psi^{\dagger}}(\tau) &=& -i\theta(\tau)\pi\rho_0 D e^{-D\tau},
	\nonumber \\ 
	g^<_{0\; \psi\psi^{\dagger}}(\tau) &=& -i\frac{\rho_0 D}{2}
	\big[e^{-D\tau}{\rm E}_1(-D\tau-i\eta)- \nonumber \\ &&
	\;\;
	e^{D\tau}{\rm E}_1(D\tau+i\eta)
	\big],
	\label{bare_g_1lead}
\end{eqnarray}
where ${\rm E}_1(z)$ is the Exponential Integral function~\cite{AS-E1}, and $\eta$ is an infinitesmal quantity that does not enter any calculation and is used only to determine which side of the branch-cut along the negative real axis in ${\rm E}_1(z)$ to take. The lesser Green function has two components, characterized by different decay behavior at long times: the real part is fast-decaying, decreasing exponentially with $D\tau$, while the imaginary component decays in a slower manner and is dominated by a $1/(D\tau)$. The bare off-diagonal Green function $g^\nu_{0\; d\psi^{\dagger}}(\tau)$ and $g^\nu_{0\; \psi d^{\dagger}}(\tau)$ are zero and the advanced Green functions are given by the relation $g^a(\tau) = \left[g^r(-\tau)\right]^*$.

At $t=0$ the hopping between the lead and the dot is turned on, and the system is driven out of equilibrium as it evolves according to the full Hamiltonian. The level acquires a finite time-dependent width which at time $t>0$ is defined by
\begin{equation}
	\Gamma_{\rm eff}(t) =
	\pi\rho_0 |\gamma - U G^<_{d\psi^{\dagger}}(t,t) |^2,
	\label{real_time_gamma_eff}
\end{equation}
where $G^<_{d\psi^{\dagger}}(t,t) = \langle \psi^{\dagger}(t)d(t) \rangle$.

It will be useful to examine first the non-interacting case $U=0$, where an exact analytical solution exists, allowing calculations of all dynamical quantities. In this case, the width of level remains time-independent at its initial value $\Gamma_0$ after the quench. However, the relevant dynamics can be extracted from calculating the expectation value of the off-diagonal matrix element $\langle \psi^{\dagger} d\rangle$ which starts from zero and reaches its equilibrium value which is given by
\begin{equation}
	\langle \psi^{\dagger} d\rangle_{\infty} = -\frac{\rho_0 \gamma}{x}
	\ln\left(\frac{1+x}{1-x}\right),
\end{equation}
where $x=\sqrt{1-4\Gamma_0/D}$, and at the wide-band-limit $D\gg \Gamma_0$ it can be approximated by $\langle \psi^{\dagger} d\rangle_{\infty}\simeq -\rho_0 \gamma\ln(D/\Gamma_0)$. The dynamics of this matrix element determine, for the interacting case $U\neq 0$, the effective width of the level $\Gamma_{\rm eff}(t)$. We relegate the presentation of the exact solution and the calculation of the dynamics to App.~\ref{Appendix:RLM_SOL}, and present here only the end result. For times $t \gg 1/D$ this matrix element is given by
\begin{eqnarray}
	\langle \psi^{\dagger}d \rangle_{(U=0)} (t)
	&\simeq& \langle \psi^{\dagger}d\rangle_{\infty} 
	+\rho_0\gamma E_1(\Gamma_0 t)
	\label{psid_nonint}
\end{eqnarray}
where ${\rm E}_1(z)$ is the exponential integral functions, and in order to get this closed analytical expression we employed the wide-band limit. This matrix element converges to its equilibrium value exponentially in time at a rate $\Gamma_0$, as this is the rate that characterizes the decay of the exponential integral function. We conclude that the dominant time-scale determining the thermalization in the non-interacting case is the width of the level $\Gamma_0$.

Turning to the interacting case, the integral equations for $\mat{G}^{\nu}(t,t')$ given by Eqs.~(\ref{quench_eqs_G_<}-\ref{quench_eqs_G_r}) take the form
\begin{eqnarray}
	\mat{G}^<(t,t') &=& \int_{0}^{t}d\tau 
	\mat{G}^<(t,\tau)\mat{\Sigma}(\tau)\mat{g}^a_0(\tau-t') + 
	\nonumber \\ &&
	\int_{0}^{t'}d\tau \mat{G}^r(t,\tau)
	\mat{\Sigma}(\tau)\mat{g}^<_0(\tau-t') +
	\nonumber \\ &&
	\mat{g}^<_0(t-t'),
	\label{quench_1_lead_G<}
\end{eqnarray}
and
\begin{eqnarray}
	\mat{G}^r(t,t') &=& 
	\int_{t'}^{t}d\tau 
	\mat{G}^r(t,\tau)\mat{\Sigma}(\tau)\mat{g}^r_0(\tau-t') +
	\nonumber \\ &&
	\mat{g}^r_0(t-t'),
	\label{quench_1_lead_Gr}
\end{eqnarray}
for $t\geq t' \geq 0$.

We have shown in Sec.~(\ref{sec:termal_eq}) that the equilibrium properties of this model are identical to those of a non-interacting Hamiltonian with $\gamma$ replaced by $\gamma_{\rm eff}$. The question arises whether the equilibrium analogy can be extended to the nonequilibrium quench: Can we obtain the time-dependent effective resonant level width from a non-interacting model where we have again replaced $\gamma$ by $\gamma_{\rm eff}$. To this end, we write a non-interacting equivalent to Eq.~(\ref{real_time_gamma_eff})
\begin{equation}
	\tilde{\Gamma}_{\rm eff}(t) =
	\pi\rho_0 |\gamma - U \langle \psi^{\dagger} d\rangle_{(U=0)} (t)|^2,
	\label{real_time_nonint_gamma}
\end{equation}
where $\langle \psi^{\dagger} d\rangle_{(U=0)} (t)$ is the exact result for the non-interacting case given in Eq.~(\ref{psid_nonint}), with the final value of $\gamma_{\rm eff}$ and $\Gamma_{\rm eff}$ replacing $\gamma$ and $\Gamma_0$.

\begin{figure}[tbf]
\centerline{
\includegraphics[width=85mm]{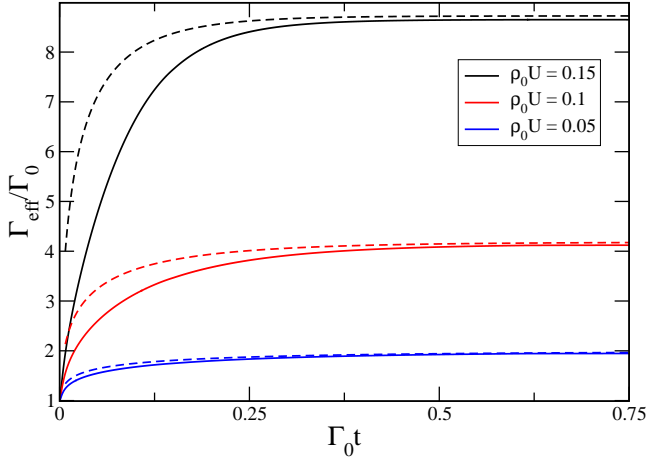}
}\vspace{0pt}
\caption{(Color online) The real-time evolution of the effective bandwidth $\Gamma_{\rm eff}$ for different values of the interaction $U$, from its initial value of $\Gamma_0$ until its equilibrium value, at zero temperature. The dashed lines are the dynamics in an effective non-interacting Hamiltonian corresponding to each interacting Hamiltonian, as defined by Eq.~(\ref{real_time_nonint_gamma}). Their values are given for the range $t\geq 5/D$. Here $\Gamma_0/D = 1.5\cdot 10^{-3}.$}
\label{Fig:quench1lead_gameff}
\end{figure}

In Fig.~(\ref{Fig:quench1lead_gameff}) we have plotted $\Gamma_{\rm eff}(t)$ for different values of $\rho_0 U$ as calculated by solving Eqs.~(\ref{quench_1_lead_G<}-\ref{quench_1_lead_Gr}) as a function of time. For comparison, we have also added $\tilde{\Gamma}_{\rm eff}(t)$ of Eq.~(\ref{real_time_nonint_gamma}), taking the dynamics from the non-interacting effective Hamiltonian as dashed lines. At long times, both $\Gamma_{\rm eff}(t)$ and $\tilde{\Gamma}_{\rm eff}(t)$ converge to the same equilibrium value, as expected. The non-interacting model is charcateried by the time scale $1/\Gamma_{\rm eff}$ from the onset. The full dynamics of the interacting model starts with the bare non-interacting value $\Gamma_{\rm eff}(t=0)=\Gamma_0$ and the fully renormalized $\Gamma_{\rm eff}(\infty)$ is dynamically built up in time, leading to the apparent slower dynamics.

Before concluding this discussion, we address here qualitatively the behavior in finite temperature. The introduction of finite temperature will effect the bare lesser Green function given in Eq.~(\ref{bare_g_1lead}) and it will read
\begin{eqnarray}
		g^<_{0\; \psi\psi^{\dagger}}(\tau) &=& \pi\rho_0 D f(-iD)e^{-D\tau} +
		\nonumber \\ &&
		\frac{2\pi i}{\beta D}\sum_{n=0}^{\infty}
		\frac{1}{1-\left(\omega_n/D\right)^2}e^{-\omega_n \tau},
\end{eqnarray}
where $f(\epsilon)$ is the Fermi-Dirac distribution and $\omega_n = (2n+1)\pi/\beta$ are the Matsubara frequencies, which here play the role of a decay rates. The slowest decaying element of the Green function will decay at a new characteristic time scale $\omega^{-1}_0 = (\pi T)^{-1}$. For temperatures smaller than $\Gamma_{\rm eff}$, the time scale $1/\Gamma_{\rm eff}$ will still characterize the system. For higher values of the temperature, this new time scale will become the dominant one and will govern the equilibration rate.

\subsection{Time evolution of the current in two-leads}

Now we extend the discussion to a two-lead setup, i.e.\ $M=2$ in Eq.~(\ref{H_general}). As in the previous section, we consider the two leads as decoupled from the level, each in equilibrium at it own chemical potential $\mu_L = -\mu_R = V/2$ for $t<0$. At time $t=0$, we connect the two leads symmetrically to the dot and shall follow the real-time evolution of the current between the leads from its initial value of zero until it reaches steady-state value calculated in Eq.~(\ref{steady_state_current_by_gamma}). At time $t$ the current will be given by Eq.~(\ref{steady_state_current_real_time}) as
\begin{equation}
	I(t) = \langle \hat{I}(t)\rangle = e\gamma
	{\rm Im}\left[
	\langle \psi^{\dagger}_R(t) d(t) \rangle - 
	\langle \psi^{\dagger}_L(t) d(t) \rangle
	\right].
	\label{eq_current_real_time_Green_f}
\end{equation}
As in the previous quench setup considered, we shall focus on the zero temperature limit $T\to 0$.

Before turning on the hopping, the dynamics of system are described by the bare Green functions, which depend only on the time difference. The bare Green functions pertaining to the level are identical to ones given for the single-lead setup in Eq.~(\ref{bare_g_dot}) while the bare Green functions of conduction electrons are slightly modified by the introduction of the chemical potential, and at $T\to 0$ they are given by
\begin{eqnarray}
	g^r_{0\; \psi_{\alpha}\psi^{\dagger}_{\alpha}}(\tau) &=& 
				-i\theta(\tau)\pi\rho_0 D e^{-D\tau},
	\nonumber \\ 
	g^<_{0\; \psi_{\alpha}\psi^{\dagger}_{\alpha}}(\tau) &=& -i\frac{\rho_0 D}{2}
	e^{-i\mu_{\alpha}\tau}
	\big[e^{-D\tau}{\rm E}_1(-D\tau-i\eta)- \nonumber \\ &&
	\;\;
	e^{D\tau}{\rm E}_1(D\tau+i\eta)
	\big],
	\label{bare_g_2lead}
\end{eqnarray}
where we have assumed a Lorentzian density of states with half-width $D$.

Following the turning on of the hopping, the finite bias between the leads results in an electrical current flowing through the level, which is manifested by $\langle\psi^{\dagger}_{\alpha}(t)d(t)\rangle$ acquiring a nonzero imaginary part. The particle-hole symmetry of the setup, described in Sec.~\ref{sec:noneq_steady_state}, guarantees that
\begin{equation}
	\langle\psi^{\dagger}_L(t)d(t)\rangle=
		\langle\psi^{\dagger}_R(t)d(t)\rangle^*,
\end{equation}
at all times.

As in the single-lead quench, the non-interacting case $U=0$ is exactly solvable in an analytical manner. At times $t\gg 1/D$, the current in the non-interacting case is given by
\begin{eqnarray}
	I_{U=0}(t) &=& 2G_0 \bar{\Gamma}_0
	\bigg\{\tan^{-1}\left(\frac{V}{2\bar{\Gamma}_0}\right)
	+ \nonumber \\ && \;\;
	{\rm Im}\left[E_1\left(\bar{\Gamma}_0 t + i\frac{V}{2}t\right)\right]
	\bigg\}.
	\label{I_non-interacting_real_time}
\end{eqnarray}
To obtain this closed expression, we employed the wide-band limit $D\gg \bar{\Gamma}_0,V$. Expanding for short times $1/D \ll t \ll {\min}\{1/\bar{\Gamma}_0,1/V\}$ the current is given by
\begin{equation}
	I_{U=0}(t) \simeq G_0\bar{\Gamma}_0 V t,
	\label{eq_current_short_times}
\end{equation}
and it grows linearly with a slope determined by $\bar{\Gamma}_0 V$. At long times the current converges to its steady-state value via exponentially decaying oscillations. The rate of convergence is $1/\bar{\Gamma}_0$ and the frequency of oscillations depends on the voltage and is $V/4\pi$.

The steady-state nonequilibrium setup is equivalent to a non-interacting model with $\bar{\Gamma}_0$ dressed to $\bar{\Gamma}_{\rm eff}$. We shall examine whether this effective non-interacting Hamiltonian can describe the real-time evolution of the system following a quench. To this end, we will use the non-interacting expression in Eq.~(\ref{I_non-interacting_real_time}) with the final dressed $\bar{\Gamma}_{\rm eff}$ replacing the bare $\bar{\Gamma}_0$.

\begin{figure}[tbf]
\centerline{
\includegraphics[width=90mm]{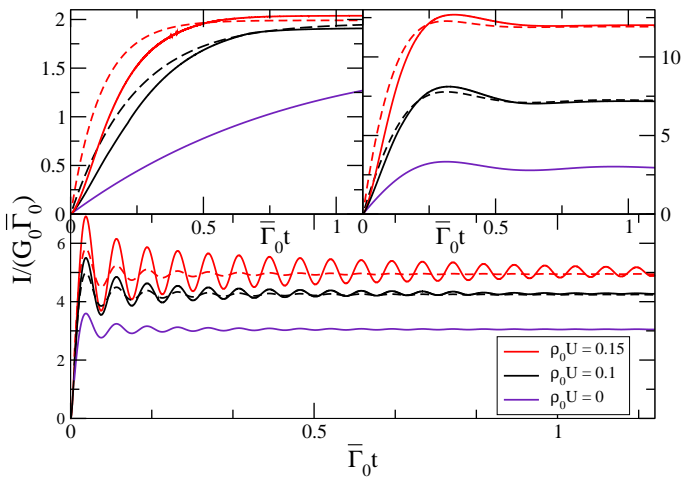}
}\vspace{0pt}
\caption{(Color online) The real-time evolution of the current, following a quench connecting the two leads to the level, for different values of the interaction $U$ and for different voltage bias. The dashed lines are the evolution of the current in an effective non-interacting Hamiltonian corresponding to the interacting one, as defined by Eq.~(\ref{I_non-interacting_real_time}). Their values are given for the range $t\geq 5/D$. Here $\bar{\Gamma}_{0}/D = 1.5 \cdot 10^{-3}$.
}
\label{Fig:quench2}
\end{figure}

%
%
%
%


We calculate numerically the current in the interacting case by solving the set of integral equations of Eqs.~(\ref{quench_eqs_G_<}-\ref{quench_eqs_G_r}), where for this setup the Green function matrices are the $3\times 3$ given in Eq.~(\ref{two_time_G_two}). In Fig.~(\ref{Fig:quench2}) we have plotted the results of these calculations, for different values of the interaction and different regimes of voltage bias. For comparison, for each interacting setup we have also plotted the time-evolution, under identical voltage bias, in an effective non-interacting setup with dressed tunneling, as given by Eq.~(\ref{I_non-interacting_real_time}).

In the low-voltage regime $V \ll \bar{\Gamma}_{\rm eff}$, the steady-state current is similar for $\rho_0 U = 0.1$ and $\rho_0 U = 0.15$. In this regime, described by Eq.~(\ref{IV_linear}), the steady-state current is independent $\bar{\Gamma}_{\rm eff}$. The value of $\bar{\Gamma}_{\rm eff}$ effects, however, the rate with which the current converges to its steady-state value. The setup with the larger value of interaction, which is characterized by a larger $\bar{\Gamma}_{\rm eff}$ at steady-state, converges faster. No oscillations are visible in the current as the frequency determined by $V$ is much slower than the rate of convergence determined by $\bar{\Gamma}_{\rm eff}$. For the non-interacting case, the bare width is smaller than the voltage bias and the steady-state current is not in the linear regime.

For intermediate values of the voltage where $V\sim \bar{\Gamma}_{\rm eff}$, the steady-state current strongly depends on the magnitude of $\bar{\Gamma}_{\rm eff}$, and therefore on the interaction. The current for the case with $\rho_0 U = 0.15$ is larger than for $\rho_0 U = 0.1$ and for the non-interacting case $U=0$. All setups show initial signs of oscillations in the currents, before arriving to the final steady-state value.

In the high-voltage regime $V \gg \bar{\Gamma}_{\rm eff}$ (but still $V\ll D$), the current shows clear oscillations before arriving to its steady-state. While the amplitude of the oscillations and the magnitude of the steady-state current depend strongly on $\bar{\Gamma}_{\rm eff}$, and thus on the interaction $U$, the frequency and phase of the oscillations depend only on the voltage, and all setups oscillate with identical frequency $V/4\pi$. The current oscillations increase in magnitude due to the interaction, and the relaxation time to steady-state is prolonged by it, which is evident from the slow relaxation of the setup with $\rho_0 U = 0.15$ compared with the noninteracting and the weaker interacting $\rho_0 U = 0.1$ setups.

Comparing the real-time evolution of the current following the quench of the interacting setups with their equivalent non-interacting effective models, we discover that while the long-time steady-state behavior is identical in both description, the dynamics are different. The interacting models are characterized by slower dynamics and stronger current oscillations. However, the frequency of oscillations, which depends on the voltage bias alone, is similar in both the interacting and the noninteracting case.

\section{Summary and Conclusions}
In this paper we studied the interacting resonant level model using a perturbative conserving approximation in the contact interaction $U$. 
We have considered a single-lead as well as a two-lead setup  assuming initially each leads in thermal equilibrium and at a fixed chemical potential. We have related the thermal equilibrium in the single-lead setup and the steady-state in the two-lead setup to the real-time evolution of the hybridization function and the time-dependent current after a quench switches on the hopping matrix element.

In Sec.~\ref{sec:termal_eq} the properties of the model in thermal equilibrium were studied, and we have benchmarked our approximation against the well-established results pertaining to that model achieved using RG techniques. The low-energy fixed point of the model describes a phase-shifted Fermi liquid, where the interaction dresses the bare width of the level $\Gamma_0$ to an effective $\Gamma_{\rm eff}$, defining the energy-scale of the model. 
In the weakly interacting regime, our approximation reproduces the equilibrium power-law renormalization of the level obtained in perturbative RG and stated in Eq.~(\ref{WC-condition-I}). This established the validity range of the our approximation.

We have calculated the steady-state current through the level in a two-lead setup at a particle-hole symmetric point as a function of the  bias and for different contact interaction strength. At low voltages the linear response regime is related to the universal regime in thermal equilibrium, and the conductance is governed by the low-energy fixed point of the IRLM at temperatures well below the characteristic energy scale $\bar\Gamma_{\rm eff}(V)$. At large bias a negative differential conductance is found and the exponent of the power-law suppression of the current has been analytically calculated. We have augmented these two analytically accessible regimes with a numerical solution for all biases to illustrate the crossover from small to large applied voltages.

The negative differential conductance reflexes the dynamically undressing of the
the strongly enhanced level width $\bar{\Gamma}_{\rm eff}(V)$ with increasing voltage: at very large voltage the original bare  level width $\bar{\Gamma}_{0}$ is recovered, and its approach is well described by a voltage dependent power-law derived in Eq.~\eqref{eq:gamma-eff-vs-bias}.
Our analytical calculations clearly reveal that both high-voltage and high-temperature serve as an effective low-energy cutoff in the self-consistency equation in a similar fashion as in RG approaches.\cite{Kennes2012}

We have extended our conserving approximation to the calculation of the fully dressed two-times Keldysh Green functions. The real-time response of the system to quantum quenches became numerically accessible for a finite contact interaction $U$. In the single-lead setup, we have followed the evolution of the width of the level from $\Gamma_0$ to the dressed $\Gamma_{\rm eff}$ after connecting the level and the lead at $t=0$. In the two-lead setup, we have calculated the evolution of the current to its steady-state value after establishing the connection between the two leads and the resonant level.

In both cases we have compared the results  at finite $U$ with the exact analytical expression derived for the dynamics in the non-interacting case. 
Although the equilibrium and steady-state properties of the model can be described by an effective non-interacting Hamiltonian with renormalized level width reflecting the Fermi-liquid fixed point in an NRG treatment, \cite{Wilson75} the real-time response after a quench cannot be fully accounted for by a simple replacement of the bare level width in the $U=0$ solution with $\Gamma_{\rm eff}$. Such a substitution lacks the time-evolution of $\Gamma_{\rm eff}(t)$  which turns out the crucial for the enhancements of the current oscillations compared to the $U=0$ solution. This enhancement of the amplitude
with increasing $U$ have also been reported in an fRG approximation to the model\cite{Kennes2012} away from particle-hole symmetry. Both approaches are well controlled and  reproduce the correct exponent of the power-law renormalization of $\Gamma_{\rm eff}(t)$ in equilibrium in the weak interaction limit. Therefore, we believe that these increasing of the oscillation
amplitude is capturing the correct physics, and are not artefacts of the approximation since the oscillations are voltage driven and already present in the exact analytical solution for $U=0$. Similar enhanced oscillations of the local level occupancy have been recently reported in quenches of the level position $\epsilon_d$ using an hybrid approach comprising the time-dependent numerical renormalization group and the time-dependent density matrix renormalization approach (td-NRG/td-DMRG). \cite{HybridTDNRG-DMRG}

Even though our approximation is restricted to small values of the interaction $U$, the model at hand displays a strong-to-weak duality, which extends also to nonequilibrium conditions. It would be interesting to compare our results with methods tailored to address strong-coupling limits, such as the hybrid td-NRG/td-DMRG.\cite{HybridTDNRG-DMRG}

\begin{acknowledgments}
The authors would like to thank Dotan Goberman for fruitful discussions and for his comments. YVA would like to thank the theoretical condensed matter group at the university of Dortmund for their kind hospitality during the course of this work.

This work was supported by the German-Israeli Foundation through grant no. 1035-36.14.

\end{acknowledgments}

\appendix

\section{Solution of the RLM}
\label{Appendix:RLM_SOL}
The non-interacting version of model, where $U=0$, is quadratic and exactly solvable. We present here an analytical solution in the wide-band limit, which allows writing the results in closed-form.

\subsection{Single lead}
For the non-interacting case, the self-energy matrix of Eq.~(\ref{self-energy-real-time-1lead}) is constant in time after the quench at $t=0$, and is given by
\begin{equation}
	\mat{\Sigma}(\tau) = \gamma 
	\left( \begin{array}{cc}
	0 & 1 \\
	1 & 0 \\
	\end{array} \right)\theta(\tau),
\end{equation}
which leads to the following equations for the dressed Green functions
\begin{eqnarray}
	G^r_{dd^{\dagger}}(t,t') &=& g^r_{dd^{\dagger}}(t,t') + \gamma 
					\int_0^{\infty}\!d\tau G^r_{d\psi^{\dagger}}(t,\tau)
						g^r_{dd^{\dagger}}(\tau-t'), \nonumber \\
	G^r_{d\psi^{\dagger}}(t,t') &=& \gamma 
					\int_0^{\infty}\!d\tau G^r_{dd^{\dagger}}(t,\tau)
						g^r_{\psi\psi^{\dagger}}(\tau-t'), \nonumber \\
	G^<_{dd^{\dagger}}(t,t') &=& g^<_{dd^{\dagger}}(t,t') + \gamma 
					\int_0^{\infty}\!d\tau \big[G^<_{d\psi^{\dagger}}(t,\tau)
						g^a_{dd^{\dagger}}(\tau-t') \nonumber \\ && \;\;
						+G^r_{d\psi^{\dagger}}(t,\tau)
						g^<_{dd^{\dagger}}(\tau-t')\big], \nonumber \\
	G^<_{d\psi^{\dagger}}(t,t') &=& \gamma 
					\int_0^{\infty}\!d\tau \big[G^<_{dd^{\dagger}}(t,\tau)
						g^a_{\psi\psi^{\dagger}}(\tau-t')
				\nonumber \\ && \;\;
				+ G^r_{dd^{\dagger}}(t,\tau)g^<_{\psi\psi^{\dagger}}(\tau-t')
				\big].
\end{eqnarray}
This set of equations can be solved in closed analytical form at zero temperature and in the wide-band limit, where the bare Green functions are given by
\begin{eqnarray}
	g^r_{dd^{\dagger}}(\tau) &=& -i\theta(\tau), 
	\label{eq_app_bare_GF_grd_wide-band} \\
	g^a_{\psi\psi^{\dagger}}(\tau) &=& -i\rho_0\theta(\tau)\delta(\tau),
	\label{eq_app_bare_GF_grp_wide-band} \\ 
	g^<_{dd^{\dagger}}(\tau) &=& \frac{1}{2},
	\label{eq_app_bare_GF_gld_wide-band} \\
	g^<_{\psi\psi^{\dagger}}(\tau) &=&
	i\rho_0\frac{1}{\tau+i\eta}.
	\label{eq_app_bare_GF_glp_wide-band}
\end{eqnarray}
Here, the $\eta$ in the lesser Green function of the conduction electrons is a small quantity, which regularizes the function for short-times, and is cutoff dependent. It is related to the bandwidth $D$ by $\eta \propto 1/D$, and we will comment later on its effects on our calculations. Focusing on the expectation value for $\langle \psi^{\dagger}d\rangle$ at time $t$ we arrive at the solution
\begin{eqnarray}
	\langle \psi^{\dagger}d\rangle(t) &=&
	G^<_{d\psi^{\dagger}}(t,t) = \rho_0\gamma 
	E_1(\Gamma_0 t) + \nonumber \\ &&
	\rho_0\gamma 
	\left[\frac{i\pi}{2}-E_1(-i\eta\Gamma_0)
	\right],
	\label{eq_app_1lead_unreg}
\end{eqnarray}
where $\Gamma_0 = \pi\rho_0|\gamma|^2$. This expression diverges for $\eta\to 0$. However, we note that the divergent term is related to the long-time expectation value $\langle \psi^{\dagger} d \rangle_{\infty}$. Expanding for small $\eta\Gamma_0$ we get
\begin{equation}
	\frac{i\pi}{2}-E_1(-i\eta\Gamma_0) = \ln\left(e^{\gamma}\eta\Gamma_0\right) +
	O(\eta\Gamma_0)
\end{equation}
with $\gamma=0.57721\ldots$ here is Euler's constant. In order to make contact with the Lorentzian density-of-states used throughout this paper, we choose the regularization $\eta \simeq (1.78 D)^{-1}$, which renders Eq.~(\ref{eq_app_1lead_unreg}) as
\begin{equation}
	\langle \psi^{\dagger}d\rangle(t) = 
	\rho_0\gamma \left[
	\ln\left(\frac{\Gamma_0}{D}\right)+E_1(\Gamma_0 t)
	\right].
\end{equation}
One should note that this expression diverges for $t\to 0$, which is also a result of the wide-band limit regularization. As the short time dynamics is governed by the fastest electronic modes, the expression is well regularized only for $t \gg 1/D$.

\subsection{Two leads}
The case of a level connected to two leads held at different chemical potentials can be generalized from the single lead. Considering a quench where at $t=0$ the hopping between the leads and the level is turned on abruptly, the Green functions for $t\geq t' \geq 0$ satisfy the following set of equations
\begin{widetext}
\begin{eqnarray}
	G^r_{dd^{\dagger}}(t,t') &=& g^r_{dd^{\dagger}}(t,t') + \gamma 
					\int_0^{\infty}\!d\tau 
						\left[ G^r_{d\psi^{\dagger}_L}(t,\tau)+
						G^r_{d\psi^{\dagger}_L}(t,\tau)\right]
						g^r_{dd^{\dagger}}(\tau-t'), \nonumber \\
	G^r_{d\psi^{\dagger}_{\alpha}}(t,t') &=& \gamma 
					\int_0^{\infty}\!d\tau G^r_{dd^{\dagger}}(t,\tau)
						g^r_{\psi_{\alpha}\psi^{\dagger}_{\alpha}}(\tau-t'),
			\nonumber \\
	G^<_{dd^{\dagger}}(t,t') &=& g^<_{dd^{\dagger}}(t,t') + \gamma 
					\int_0^{\infty}\!d\tau 
					\left[G^<_{d\psi^{\dagger}_L}(t,\tau)+
					G^<_{d\psi^{\dagger}_R}(t,\tau)\right]
						g^a_{dd^{\dagger}}(\tau-t') + \nonumber \\ && \;\;
						+\int_0^{\infty}\!d\tau 
						\left[G^r_{d\psi^{\dagger}_L}(t,\tau)+
					G^r_{d\psi^{\dagger}_R}(t,\tau)\right]
						g^<_{dd^{\dagger}}(\tau-t'), \nonumber \\
	G^<_{d\psi^{\dagger}_\alpha}(t,t') &=& \gamma 
					\int_0^{\infty}\!d\tau \left[G^<_{dd^{\dagger}}(t,\tau)
						g^a_{\psi_\alpha\psi^{\dagger}_\alpha}(\tau-t')
				+ G^r_{dd^{\dagger}}(t,\tau)
				g^<_{\psi_\alpha\psi^{\dagger}_\alpha}(\tau-t')
				\right],
\end{eqnarray}
\end{widetext}
with $\alpha=L,R$ the different leads. The current is given by Eq.~(\ref{eq_current_real_time_Green_f}), and for a symmetric setup $\mu_L=-\mu_R=V/2$, at resonance, it suffices to calculate the imaginary part of $\langle\psi^{\dagger}_L d \rangle$ at time $t$.

Restricting attention to zero temperature, and employing the wide-band limit, the current can be calculated in closed analytical form. The bare Green functions for the level are the same as in the singlel lead setup and are given by Eqs~(\ref{eq_app_bare_GF_grd_wide-band}) and (\ref{eq_app_bare_GF_gld_wide-band}). This also holds for the bare retarded and advanced functions pertaining to the leads, which are identical for both leads and are still given by Eq.~(\ref{eq_app_bare_GF_grp_wide-band}). Introducing chemical potential to the leads changes only the bare lesser Green function of the lead $\alpha$, which reads
\begin{equation}
	g^<_{\psi_\alpha\psi^{\dagger}_\alpha}(\tau) =
	i\rho_0e^{-i\mu_\alpha\tau}\frac{1}{\tau+i\eta},
\end{equation}
with $\mu_\alpha$ the chemical potential.

Solving the equations for $t=t'>0$, assuming a symmetric setup at resonance, we find that the current is given at this limit by
\begin{eqnarray}
	I(t) &=& 2G_0\bar{\Gamma}_0\bigg[
	\tan^{-1}
	\left(\frac{V}{2\bar{\Gamma}_0}
	\right)+
	\nonumber \\ &&
	{\rm Im}
	\left\{E_1\left(\bar{\Gamma}_0 t +
	i\frac{V}{2}t\right)\right\}
	\bigg],
\end{eqnarray}
with $\bar{\Gamma}_0 = 2\pi\rho_0|\gamma|^2$. Here the regularization of $\eta$ does not play a role, as only the real part of $\langle \psi^{\dagger}(t)d(t)\rangle$ diverges for $\eta\to 0$, while the current depends solely on the imaginary part.

\end{document}